\documentclass[conference]{IEEEtran}
\IEEEoverridecommandlockouts
\usepackage{cite}
\usepackage{amsthm}
\usepackage{bbm}
\usepackage{amsmath,amssymb,amsfonts}
\newtheorem{theorem}{Theorem}
\newtheorem{example}{Example}
\newtheorem{definition}{Definition}
\newtheorem{lemma}{Lemma}

\newtheorem{corollary}{Corollary}
\newtheorem{remark}{Remark}

\usepackage{algorithmic}
\usepackage{graphicx}
\usepackage{textcomp}
\usepackage{xcolor}
\usepackage{comment}
\usepackage{hyperref}

\newcommand{\Nf}{\mathcal N_{\mathrm{full}}}
\def\BibTeX{{\rm B\kern-.05em{\sc i\kern-.025em b}\kern-.08em
    T\kern-.1667em\lower.7ex\hbox{E}\kern-.125emX}}
\begin{document}

\title{Distributed Detection Under a Covertness Constraint
}

 \author{
    \IEEEauthorblockN{Ismaila Salihou Adamou, Mich\`ele Wigger}
        \IEEEauthorblockA{  L2S, CentraleSupelec, Uniersite Paris-Saclay, 3, rue Joliot Curie
91190 Gif-sur-Yvette
France 
    \\\{ismaila.salihou-adamou, michele.wigger\}@centralesupelec.fr}
}

\maketitle
\sloppy
\allowdisplaybreaks

\begin{abstract}
We study distributed binary hypothesis testing over discrete memoryless channels (DMCs) under a covertness constraint, which requires that an external warden be unable to detect communication between the remote sensor and a decision center. We determine the Stein-exponent for two distinct scenarios: one   where covertness is imposed  under the null hypothesis only, and the second where covertness is imposed  under both hypotheses.  Our results show that when the DMC is partially-connected (i.e., when certain output symbols can only be induced by some of the inputs), the  Stein-exponent is the same under  both scenarios  and moreover coincides with  Shalaby and Papamarcou's Stein-exponent for distributed hypothesis testing with a sublinear number of noise-free communication bits.  In particular, this exponent does not depend on the specific transition probabilities of the DMC. When the DMC is fully-connected, the Stein-exponent is generally smaller than for partially-connected DMCs, differs in the two scenarios, and depends on the specific transition law of the DMC. While our converse results hold for arbitrary long shared secret keys between transmitter and legitimate receiver, most of our achievability results do not rely on a secret key at all. Only the coding scheme for partially-connected DMCs and only in the scenario where covertness is imposed under both hypotheses requires a secret key, whose length is slightly larger than the logarithm of the observation length.
\end{abstract}


\section{Introduction}
\label{sec:introduction}

\begin{figure}[t!]
\centering
\includegraphics[scale=0.84]{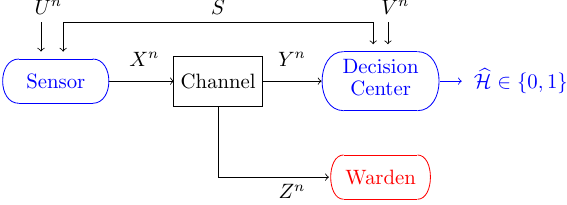}
\caption{Distributed hypothesis testing with an external warden.}
\vspace{-3mm}
\label{fig:system_model}
\end{figure}

In distributed hypothesis testing, two distant terminals, a remote sensor and a decision center, observe correlated sequences $U^n=(U_1,\ldots,U_n)$ and $V^n=(V_1,\ldots,V_n)$, respectively, see Figure~\ref{fig:system_model} (without the external warden). Under each hypothesis, $\mathcal{H}=0$ or $\mathcal{H}=1$, the pairs $(U_i,V_i), (i\in{1,\ldots,n})$, are independent and identically distributed (i.i.d.) according to a joint distribution that depends on the hypothesis. The goal of the decision center is to identify the true hypothesis based on its own observations and communication from the sensor. The performance of the decision center is characterized by two error probabilities \cite{lehmann2005testing}: the Type-I error probability of declaring $\hat{\mathcal{H}}=1$ when $\mathcal{H}=0$, denoted by $\alpha_n$, and the Type-II error probability of declaring $\hat{\mathcal{H}}=0$ when $\mathcal{H}=1$, denoted by $\beta_n$.

An important line of research in Information Theory aims at characterizing the largest exponential decay rate of the Type-II error probability $\beta_n$ under the constraint that the  asymptotic Type-I error probability $\alpha_n$ does not exceed a prescribed threshold $\epsilon\in[0,1)$ \cite{Han1987,ahlswede1986hypothesis,shimokawa1994error,Watanabe_DHT,kochman2025improved,Kochman,Michele, wen2026exact,wu2026distributed}. This largest exponent is commonly referred to as the \emph{Stein-exponent}. It has been extensively studied in centralized settings and in distributed settings where communication from the sensor to the decision center takes place over a noiseless link. Even in the latter case, a general characterization of the Stein-exponent remains an open problem, with notable exceptions for certain source distributions \cite{ahlswede1986hypothesis,Rahman2012,wen2026exact,wu2026distributed} and when the number of transmitted noise-free bits grows sublinearly with the observation length $n$ \cite{Shalaby}.

In a recent work \cite{bounhar2026distributed}, the results in \cite{Shalaby} were extended  to settings where the sensor communicates with the decision center over a discrete memoryless channel (DMC). %Such settings arise naturally in practical applications where sensors are often subject to stringent energy constraints. 
The results in \cite{ITW} established a dichotomy in the Stein-exponent depending on the connectivity of the DMC. When the DMC is fully-connected, i.e., every input symbol can induce every output symbol, the Stein-exponent coincides with the local exponent at the decision center. Thus, communication from the sensor does not improve the Stein-exponent. In contrast, when the DMC is partially-connected, i.e., certain output symbols cannot be induced by some channel inputs, the same exponent is achievable as over a noiseless link \cite{Shalaby}. This result remains valid when the sensor can use the DMC $n$ times, where $n$ denotes the observation length, but is subject to a stringent sublinear cost constraint. The results however change when only a relaxed expected sublinear cost constraint is imposed \cite{bounhar2026distributed}. For partially-connected DMCs, the Stein-exponent still coincides with the noiseless-link exponent in \cite{Shalaby}. For fully-connected DMCs, however, the relaxed expected cost constraint allows an improvement over the local Stein-exponent at the decision center.

In the present work, the sensor again communicates with the decision center over $n$ uses of a DMC, but communication is subject to a \emph{covertness constraint}, see Figure~\ref{fig:system_model}. Specifically, an external warden observes the output of another DMC and should be unable to distinguish the induced output distribution from that corresponding to the absence of communication. Depending on the practical motivation, covertness may be required only under the null hypothesis, corresponding to a non-alert situation, or under both hypotheses. In this paper, we consider both scenarios. As is common in covert communication, we allow the sensor and the decision center to share a uniform secret key that is unknown to the warden.

Under standard assumptions on the warden's channel \cite{Bloch2016, Wang2016}, the covertness constraint implies an expected sublinear cost constraint on the channel inputs. As we show in this paper, the optimal Stein-exponents under covertness constraints coincide with those established in \cite{bounhar2026distributed} under an expected sublinear cost constraint. Thus, although covertness imposes a stronger restriction on communication, it does not reduce the optimal Stein-exponent compared to an expected sublinear cost constraint.

Despite this coincidence, both the achievability and converse proofs require modifications. In particular, the achievability schemes must be adapted to satisfy the stronger covertness constraints, while the converse proofs must account for the secret key shared between the sensor and the decision center, whose length is unrestricted. In this context, we also identify a gap in one of the converse proofs in \cite{bounhar2026distributed}. This gap can be addressed using the converse techniques developed in the present paper, and a corrected proof of the corresponding result is provided in \cite{corrected_proof}.

Interestingly, most of our achievability results do not require a secret key. An exception arises for partially-connected channels when covertness is imposed under both hypotheses. In this setting, our proposed scheme requires a secret key whose length grows logarithmically with the observation length. Moreover, we show that the optimal Stein-exponent cannot be achieved without a secret key in this setting.

To summarize, our results show that for partially-connected DMCs, the optimal Stein-exponent coincides with Shalaby and Papamarcou's exponent for distributed hypothesis testing with a sublinear number of noise-free communication bits \cite{Shalaby}, irrespective of whether covertness is imposed only under the null hypothesis or under both hypotheses.  The exponent thus does not depend on the specific transition probabilities of the DMC.
For fully-connected DMCs, when  covertness is imposed only under the null hypothesis, the Stein-exponent is determined by the minimum of two quantities: the Shalaby-Papamarcou noiseless-link exponent and an exponent depending on the local observations and the DMC's detection exponent. When covertness is imposed under both hypotheses, an additional third exponent appears in the minimization, which generally decreases the Stein-exponent. In all cases, the optimal Stein-exponent does not depend on the prescribed Type-I error threshold $\epsilon$.

This paper's setup has previously been considered in \cite{bounhar2024covert}, where it was shown that the noiseless-link exponent provides an upper bound on the Stein-exponent under a covertness constraint. The converse in \cite{bounhar2024covert}, however, assumes that the secret-key length grows sublinearly with the observation length $n$. In the present work, we extend this converse proof to arbitrary secret-key lengths. Moreover, the work in \cite{bounhar2024covert} did not establish any achievability result improving upon the local exponent, whereas our results demonstrate that such an improvement is possible.

It is also worth mentioning the related works in \cite{sreekumar2020privacy,Mhanna,TACI_HT,Faour}, where privacy constraints are imposed on distributed hypothesis testing. In these settings, a warden eavesdropping on the noiseless communication link from the sensor to the decision center should be unable to infer certain information about the sensor's observations. For example, \cite{TACI_HT} characterized the optimal Stein-exponent under a privacy constraint for the testing-against-conditional-independence setting when the Type-I error probability is required to vanish. More recently, \cite{Faour} extended this result to arbitrary constraints $\epsilon\in[0,1)$ on the Type-I error probability and showed that the Stein-exponent can depend on $\epsilon$.

\textbf{\textit{Notation:}} We mostly follow standard notation. Random variables are denoted by uppercase letters (e.g., $X$), while their realizations are denoted by lowercase letters (e.g., $x$). We abbreviate $(x_1,\ldots,x_n)$ by $x^n$ and $(x_{t+1},\ldots,x_n)$ by $x_{t+1}^n$. The Hamming weight and Hamming distance are denoted by $w_{\textnormal{H}}(\cdot)$ and $d_{\textnormal{H}}(\cdot,\cdot)$, respectively. We further abbreviate \emph{independent and identically distributed} as \emph{i.i.d.} and \emph{probability mass function} as \emph{pmf}.

We denote by $\pi_{x^n y^n}$ the joint type of the sequences $(x^n,y^n)$, defined as
\begin{equation}
\pi_{x^n y^n}(a,b) \triangleq \frac{n_{x^n,y^n}(a,b)}{n},
\end{equation}
where $n_{x^n,y^n}(a,b)$ is the number of occurrences of the pair $(a,b)$ in $(x^n,y^n)$. We use $\mathcal{T}_{\mu}^{(n)}(P_{XY})$ to denote the jointly strongly typical set as defined in \cite[Definition~2.9]{Csiszarbook}.

We denote mutual information, Kullback--Leibler (KL) divergence,
and entropy by $I(\cdot;\cdot)$, $D(\cdot\|\cdot)$,
and $H(\cdot)$, respectively.
The binary entropy function and binary KL divergence are defined as
\begin{IEEEeqnarray}{rCl}
H_{\mathrm b}(p)&\triangleq&-p\log p-(1-p)\log(1-p),
\end{IEEEeqnarray}
and
\begin{IEEEeqnarray}{rCl}
D_{\mathrm b}(p\|q)&\triangleq&
p\log\frac{p}{q}+(1-p)\log\frac{1-p}{1-q},
\end{IEEEeqnarray}
respectively.
Throughout this paper, all logarithms are natural (base \(e\)), unless stated otherwise.
 %and $\mathcal{T}_{0}^{(n)}(P_{XY})$ to denote the set of all sequence pairs of constant joint type $P_{XY}$.

We use Landau notation: $o(1)$  denotes any function that tends to zero as some parameter $n \to \infty$, and $O(f(n))$ denotes any function that grows in the order of $f(n)$.

%Throughout this manuscript, $\{\mu_n\}_{n=1}^{\infty}$ denotes a sequence of positive numbers satisfying\footnote{Condition~\eqref{eq:sq} ensures that the probability of the strongly typical set $\mathcal{T}_{\mu_n}^{(n)}(P_{XY})$ under $P_{XY}^{\otimes n}$ converges to one as $n \to \infty$; see \cite[Remark following Lemma~2.12]{Csiszarbook}.}
%\begin{subequations}\label{eq:mu_seq}
%\begin{IEEEeqnarray}{rCl}
%\lim_{n \to \infty} \mu_n &=& 0, \label{eq:M}\\
%\lim_{n \to \infty} n \mu_n^2 &=& \infty. \label{eq:sq}
%\end{IEEEeqnarray}
%\end{subequations}

\section{Problem Setup} 
\label{sec:pb_setup}
Consider the hypothesis testing problem in Figure~\ref{fig:system_model}, where the sensor observes a sequence $U^n$ and a secret key $S$ and communicates to a decision center, which also knows the secret key $S$ in addition to its local observations $V^n$.  The secret key $S$ is uniformly distributed over a finite set $\mathcal K$ independent of $(U^n,V^n)$ under both hypotheses, and samples $U_i$ and $V_i$, for $i=1,\ldots, n$, take value in given finite alphabets $\mathcal{U}$ and $\mathcal{V}$, respectively.
Under the null hypothesis
\begin{equation}
    \mathcal{H}=0 : (U^n, V^n) \  \textnormal{i.i.d.}  \ P_{UV}
\end{equation}
whereas under the alternative hypothesis
\begin{equation}
    \mathcal{H}=1 : (U^n, V^n) \  \textnormal{i.i.d.}  \ Q_{UV}
\end{equation}
for given pmfs $P_{UV}$ and $Q_{UV}$ over $\mathcal{U}\times \mathcal{V}$. 
Similarly to \cite{Shalaby} it is assumed that %$P_{UV}(u,v)=0$ whenever 
$Q_{UV}(u,v)>0$ for all pairs $(u,v)\in\mathcal{U}\times \mathcal{V}$.

The sensor sends an input sequence
\begin{equation}
    X^n = f^{(n)}(U^n, S) \in \mathcal{X}^n
\end{equation}
over the channel, where $f^{(n)}$ is a chosen encoding function on appropriate domains. The decision center observes the corresponding outputs $Y^n$ of the discrete memoryless channel (DMC) $\Gamma_{YZ|X}$ and the warden the outputs $Z^n$. Output symbols $Y_i$ and $Z_i$, for $i\in\{1,\ldots, n\}$, take values in the finite alphabets $\mathcal{Y}$ and $\mathcal{Z}$, respectively. 

It is assumed that the input alphabet $\mathcal{X}$ contains the 0 symbol. Define $\tilde{\mathcal{X}}$ as the  set of all non-zero input symbols
\begin{equation} 
\tilde{\mathcal{X}} \triangleq \mathcal{X} \backslash \{0\}.
\end{equation}
Covertness  imposes that  the pmf of the warden's output distribution  
\begin{equation}
\label{eq:P_z_H_0}   P_{Z^n|\mathcal{H} = 0} = \frac{1}{|\mathcal{K}|} \sum_{s \in \mathcal{K}} \sum_{u^n \in \mathcal{U}^n} P_{U}^{\otimes n} (u^n) \Gamma_{Z|X}^{\otimes n} (\cdot|f^{(n)}(u^n, s)),
\end{equation}
under $\mathcal{H}=0$, and  
\begin{equation}
\label{eq:P_z_H_1}   P_{Z^n|\mathcal{H} = 1} = \frac{1}{|\mathcal{K}|} \sum_{s \in \mathcal{K}} \sum_{u^n \in \mathcal{U}^n} Q_{U}^{\otimes n} (u^n) \Gamma_{Z|X}^{\otimes n} (\cdot|f^{(n)}(u^n, s)),
\end{equation}
under $\mathcal{H}=1$
be close to the output distribution $\Gamma_{Z|X=0}^{\otimes n}$ induced by the all-zero input sequence. Specifically, covertness is measured 
by the Kullback-Leibler divergence:
\begin{equation}
    d_{n,H}\triangleq D(P_{Z^n|\mathcal{H} = H} || \Gamma_{Z|X=0}^{\otimes n}) \ \ \ \  H \in \{0, 1\}
\end{equation}
which is required to vanish as $n\to\infty$ under each hypothesis for which covertness is imposed.

Following 
standard convention in covert communication, we require that the DMC $\Gamma_{YZ|X}$ satisfies the following  conditions: 
{\begin{subequations}
\label{eq:prelude_channel_conditions_p2p}
    \begin{IEEEeqnarray}{rCl}
  \sum_{x \in \mathcal{X} \backslash \{0\}} \psi (x)  \Gamma_{Z|X}(\cdot| x)& \neq & \Gamma_{Z|X}(\cdot| 0) ,\quad \forall \  \psi(\cdot),\label{eq:a14} \IEEEeqnarraynumspace \\
  %\sum_{x \in \mathcal{X} \backslash \{0\}} \psi(x)  \Gamma(\cdot| x)& \neq & \Gamma(\cdot| 0),\quad \forall \psi(\cdot), \label{eq:a12}\\
    \textnormal{Supp} \left(\Gamma_{Z|X}(\cdot| x )\right) &\subseteq&      \textnormal{Supp} \left(\Gamma_{Z|X}(\cdot| 0)\right),\quad \forall x \in \mathcal{X},\label{eq:a13} \IEEEeqnarraynumspace \\
        \textnormal{Supp} \left(\Gamma_{Y|X}(\cdot| x) \right) &\subseteq&      \textnormal{Supp} \left(\Gamma_{Y|X}(\cdot| 0)\right), \quad \forall x \in \mathcal{X}, \label{eq:a11} \IEEEeqnarraynumspace
    \end{IEEEeqnarray}
\end{subequations}
where in the above, $\psi(\cdot)$ indicates any pmf over $\mathcal{X}\backslash \{0\}$.}

Based on the received sequence $Y^n $,  its observations $V^n$, and the secret key $S$, the decision center produces a guess of the hypothesis:
\begin{equation}\label{eq:guess}
\hat{\mathcal{H}} =	g^{(n)}(V^n,Y^n, S) \in\{0,1\}.
\end{equation}

The goal is to design encoding and decision functions such
that the Type-I  error probability 
\begin{equation}
\alpha_n \triangleq \Pr \left[ \hat{\mathcal{H}}=1 |\mathcal{H}=0 \right] 
\end{equation} 
stays below given threshold $\epsilon \geq 0$ and the Type-II 
error probability
\begin{equation}
\beta_n \triangleq  \Pr \left[ \hat{\mathcal{H}}=0 |\mathcal{H}=1 \right]
\end{equation}
decays exponentially fast with the largest possible exponent,
subject to the covertness constraint
\begin{equation}\label{eq:cov}
\lim_{n\to\infty}d_{n,H}=0,
\end{equation}
for each hypothesis $H$ under which covertness is imposed.
We will consider two scenarios: a first one where covertness is only required under $H=0$,
and a second one where  covertness is required under both $H=0$ and $H=1$.

 \begin{definition}\label{def:ach} Given  $\epsilon\in[0,1)$, a miss-detection error exponent $\theta>0$ is called $\epsilon$-achievable \emph{under a covertness constraint} if there exists a  sequence of encoding and decision functions   $\{(f^{(n)},g^{(n)})\}_{n=1}^\infty$  satisfying \eqref{eq:cov} and 
\begin{subequations}\label{eq:criteria2}
	\begin{IEEEeqnarray}{rCl}
	\varlimsup_{n\to \infty} \alpha_n %\Pr \left[ \hat{\mathcal{H}}=1 |\mathcal{H}=0 \right] 
    &\leq&  \epsilon \label{eq:P12}\\
\varliminf_{n\to \infty}- \frac{1}{n} \log % \Pr \left[ \hat{\mathcal{H}}=0 |\mathcal{H}=1 \right] 
\beta_n &\geq&  \theta \label{eq:det_exp2}. %,\\
%\varlimsup_{n\to \infty} \frac{\log(K)}{\sqrt{n \delta_{n,H}}} &\leq& R_s. \label{eq:key_rate}
	\end{IEEEeqnarray} 
	%and \begin{equation}\label{eq:cov}
%	\lim_{n\to \infty} d_{n,0} =0.
%	\end{equation}
\end{subequations}
The supremum over all $\epsilon$-achievable  miss-detection error exponents $\theta$ is denoted by %$\theta_{\textnormal{covert}}^\star(\epsilon)$ and called  \emph{covert Stein-exponent}.
% We distinguish between two covertness scenarios. We denote by 
$\theta_{\textnormal{covert},\mathcal H=0}^{*}(\epsilon)$ when covertness is imposed only under $\mathcal H=0$, and by $\theta_{\textnormal{covert,both}}^{*}(\epsilon)$ when covertness is imposed  under both hypotheses. %the optimal Stein-exponent when covertness is imposed under both $\mathcal H=0$ and $\mathcal H=1$. 

 \end{definition}

As we shall see, the Stein-exponent will depend  on source pmfs $P_{UV}$ and $Q_{UV}$ as well as on the connectivity of the channel $\Gamma_{YZ|X}$. We therefore define the following  two classes of channels.
\begin{definition} \label{def:class_chan}
Let $\mathcal{N}$ denote the set of DMCs $\Gamma_{YZ|X}$ satisfying Conditions~\eqref{eq:prelude_channel_conditions_p2p}. We define $\mathcal N_{\mathrm{partial}}\subseteq\mathcal N$ as the class of channels for which there exists an input $\hat{x}\in\mathcal{X} \setminus \{0\}$ and an output $y^*\in \mathcal{Y}$ satisfying the two conditions:
\begin{subequations}\label{eq:channel_cond0}
\begin{IEEEeqnarray}{rCl}
\Gamma_{Y|X}(y^*|0) &> &0 \\
\Gamma_{Y|X}(y^*|\hat{x}) &= &0.
\end{IEEEeqnarray}
\end{subequations}
We further define $\mathcal N_{\mathrm{full}}\triangleq\mathcal N\setminus\mathcal N_{\mathrm{partial}}$.
\end{definition}

\begin{remark}
By the definition of $\mathcal{N}$, see Conditions~\eqref{eq:prelude_channel_conditions_p2p}, any channel $\Gamma_{YZ|X}\in \mathcal N_{\mathrm{full}}$ is fully-connected, i.e., 
\begin{eqnarray}
\Gamma_{Y|X}(y|x) >0, \quad \forall (x,y)\in\mathcal{X}\times \mathcal{Y}.
\end{eqnarray}
In fact, under Conditions~\eqref{eq:prelude_channel_conditions_p2p}, if
\begin{equation}
\Gamma_{Y|X}(y^*|x')>0 
\qquad\textnormal{and} \quad 
\Gamma_{Y|X}(y^*|\hat{x})=0
\end{equation}
for some $(y^*,\hat{x}, x') \in \mathcal{Y}\times \mathcal{X} \times \mathcal{X}$  then also $\Gamma_{Y|X}(y^*|0)>0$, because 
%directly imply that $x'=0$, i.e.,  \eqref{eq:channel_cond0}, or otherwise lead to a contradiction.
%However, under the standing assumption \eqref{eq:a13}, commonly adopted in
%covert communication models, this condition is equivalent to
%\eqref{eq:channel_cond0}.
%Indeed, $\Gamma_{Y|X}(y^*|x')>0$ implies
%$y^*\in\operatorname{Supp}(\Gamma_{Y|X}(\cdot|x'))$, and hence,
by assumption \eqref{eq:a11},
$\operatorname{Supp}(\Gamma_{Y|X}(\cdot|x'))\subseteq \operatorname{Supp}(\Gamma_{Y|X}(\cdot|0))$.

%The only remaining class of partially-connected channels not covered by
%the theorem are those for which some nonzero input can produce an output
%that cannot be produced by the zero input, i.e., there exist
%$x\neq 0$ and $y^*\in\mathcal{Y}$ such that
%\[
%\Gamma_{Y|X}(y^*|x)>0,\qquad
%\Gamma_{Y|X}(y^*|0)=0.
%\]
%However, such channels violate the support inclusion condition \eqref{eq:a13}   and are therefore excluded from the considered channel model.
\end{remark}

%%%%%%%
\section{Results}\label{sec:result}

Define the following exponents

\begin{subequations}
\begin{IEEEeqnarray}{rCl}
E_1 & \triangleq &  \min_{\substack{\pi_{UV} \colon \\
\pi_U=P_U\\
\pi_V=P_V}} D( \pi_{UV} \| Q_{UV}) \\
E_2 & \triangleq &D( P_{V} \| Q_{V}) + \max_{x_1 \in\tilde{\mathcal{X}}}  D( \Gamma_{Y|X=0}\| \Gamma_{Y|X=x_1}) \IEEEeqnarraynumspace \label{eq:max4b}\\[1.2ex]
E_3 & \triangleq &  \min_{\substack{\pi_{UV} \colon \\
\pi_U=Q_U\\
\pi_V=P_V}} D( \pi_{UV} \| Q_{UV}) , 
\end{IEEEeqnarray} 
\end{subequations}
where notice that the difference between $E_1$ and $E_3$ only lies in the  minimization constraints $\pi_U=P_U$ or $\pi_U=Q_U$. Also, notice that for channels $\Gamma_{YZ|X}\in\Nf$ the divergence term $D(\Gamma_{Y|X=0}\| \Gamma_{Y|X=x_1})<\infty$ for all $x_1\in\tilde{\mathcal{X}}$ and thus $E_2$ is finite for these channels.

We can now state our main theorems.
The first Theorem %is when we  
concerns the scenario with  covertness constraint  under $\mathcal{H}=0$ only, i.e., we require
\begin{equation} \label{eq:cov_h0}
\lim_{n\to \infty}d_{n,0}=0.
\end{equation} 
\begin{theorem}[Covertness under $\mathcal{H}=0$ only]\label{sec:theo1}
	Fix $\epsilon \in [0,1)$.
\begin{enumerate}
\item If $\Gamma_{YZ|X} \in \mathcal N_{\mathrm{partial}}$ then:
\begin{IEEEeqnarray}{rCl}\label{eq:Ia}
\theta_{\textnormal{covert}, \mathcal{H}=0}^{*}(\epsilon) &=&  E_1.
\end{IEEEeqnarray}
\item If $\Gamma_{YZ|X}\in \Nf$, then:
\begin{IEEEeqnarray}{rCl}
\theta_{\textnormal{covert}, \mathcal{H}=0}^{*}(\epsilon) = \min\big\{   E_1, E_2 \big \}\IEEEeqnarraynumspace\label{eq:Ib}. 
\end{IEEEeqnarray}
\end{enumerate}

Our achievability results do not require that the sensor and decision center use the shared secret key~$S$.

%The covertness constraint $d_{n,0}$ can be made to vanish exponentially fast in $n$ for the result in \eqref{eq:Ib} and arbitrary close to exponentially fast in $n$ for the achievability result in \eqref{eq:Ia}.
\end{theorem}
\begin{IEEEproof}
    Achievability results are proved in Section~\ref{sec:proof_achiev} and converses in Section~\ref{sec:proof_conv}.
\end{IEEEproof}
\begin{remark}
Exponent $E_1$ is also the Stein-exponent in a distributed setup without warden where the sensor can send a sublinear (in $n$) number of noise-free bits to the decision center \cite{Shalaby}. For channels $P_{YZ|X}\in \mathcal N_{\mathrm{partial}}$, the covertness constraint \eqref{eq:cov} thus has the same impact as limiting communication to a  sublinear (in $n$) number of noise-free bits (without covertness constraint).
\end{remark}

Our second Theorem assumes a covertness constraint  under both hypotheses:
\begin{equation}
\lim_{n\to \infty}d_{n,H}=0, \textnormal{ for }H\in\{0,1\}.
\end{equation} 
  
\begin{theorem}[Covertness under Both Hypotheses]
\label{sec:theo2}
	Fix $\epsilon \in [0,1)$.
\begin{enumerate}
\item If $P_{YZ|X} \in \mathcal N_{\mathrm{partial}}$ then:
\begin{IEEEeqnarray}{rCl}
\theta_{\textnormal{covert,both}}^{*}(\epsilon) &=&    E_1.\label{eq:bothpart}
\end{IEEEeqnarray}
\item If $P_{YZ|X}\in \Nf$, then:
\begin{IEEEeqnarray}{rCl}
\theta_{\textnormal{covert,both}}^{*}(\epsilon) &=& \min\big\{   E_1, E_2, E_3 \big \} . \label{eq:bothfull}\IEEEeqnarraynumspace
\end{IEEEeqnarray}

\end{enumerate}
Achievability of \eqref{eq:bothfull} can be proved without using the shared secret key $S$. In our achievability proof of \eqref{eq:bothpart} the sensor and the decision center need a secret key $S$ consisting of $O(m_n \log n)$ bits, for an arbitrary  function $m_n$ that grows with $n$.  \end{theorem}

    \begin{IEEEproof}
    Achievability results are proved in Section~\ref{sec:proof_achiev} and converses in Section~\ref{sec:proof_conv}.
\end{IEEEproof}

We notice that unlike for partially-connected channels, for fully-connected channels the optimum Stein-exponent is decreased when covertness is required under both hypotheses and not only under $\mathcal{H}=0$. 

%Notice that Theorems 1 and 2 differ only for channels in $\mathcal{N}_{\textnormal{full}}$.

\begin{corollary}
[Improvement over the Local Test]
\label{sec:improve_local}
Consider the pmf 
\begin{equation} 
T_{U}(u)\triangleq \sum_{v} P_V(v)\,Q_{U|V}(u\mid v) ,\quad u \in \mathcal{U},
\end{equation}
and assume a channel $\Gamma_{YZ|X}\in\mathcal{N}$ for which %\footnote{Notice that the Assumptions~\eqref{eq:prelude_channel_conditions_p2p} imply that $
%D(\Gamma_{Y|X=0}\|\Gamma_{Y|X=x_1})>0$ for some $x_1 \neq 0$.} 
\begin{equation} 
\max_{x_1\in \tilde{\mathcal{X}}}D(\Gamma_{Y|X=0}\| \Gamma_{Y|X=x_1})>0.\label{eq:pos} 
\end{equation}
If $T_U \neq P_U$ then for every $\epsilon\in[0,1)$:
\begin{IEEEeqnarray}{rCl}
\theta_{\textnormal{covert}, \mathcal{H}=0}^{*}(\epsilon)
&>& D(P_V\|Q_V),
\end{IEEEeqnarray}
If $T_U \notin \{ P_U, Q_U\}$, then   for every $\epsilon\in[0,1)$ also:
\begin{IEEEeqnarray}{rCl}
\theta_{\textnormal{covert,both}}^{*}(\epsilon)
&>& D(P_V\|Q_V).
\end{IEEEeqnarray}
\end{corollary}

\begin{IEEEproof}
Notice that for any $\pi_{UV}$ with marginal $\pi_V=P_V$:
\begin{IEEEeqnarray}{rCl}
\lefteqn{D\big(\pi_{UV}\big\|Q_{UV}\big)}
\; \nonumber\\
&=& D\big(P_V\big\|Q_V\big)
+\sum_{v} P_V(v)\,
D\big(\pi_{U|V}(\cdot|v)\big\|
Q_{U|V}(\cdot|v)\big).
\IEEEeqnarraynumspace
\end{IEEEeqnarray}

The second summand is strictly positive unless
$\pi_{U|V}(\cdot|v)=Q_{U|V}(\cdot|v)$
for all $v$ with $P_V(v)>0$. Thus, to deduce that
$E_1>D(P_V\|Q_V)$ or
$E_3>D(P_V\|Q_V)$,
it suffices to show that
$\pi_{U|V}=Q_{U|V}$
is not a permissible choice in the corresponding
minimization.

Notice that the choice $\pi_{U|V}=Q_{U|V}$ induces the marginal
\[
\pi_U(u)
=
\sum_v P_V(v)Q_{U|V}(u|v)
=
T_U(u).
\]
Therefore, $\pi_{U|V}=Q_{U|V}$ is not a permissible choice  in the
minimization of $E_1$ whenever $T_U\neq P_U$,
nor in the minimization of $E_3$
when $T_U\neq Q_U$.

Consequently, if $T_U\neq P_U$:
\begin{equation}
E_1>D(P_V\|Q_V),
\end{equation}
and if $T_U\neq Q_U$:
\begin{equation}
E_3>D(P_V\|Q_V).
\end{equation}

To conclude the proof, notice  that Condition~\eqref{eq:pos} implies 
\begin{equation}
E_2>D(P_V\|Q_V).
\end{equation}

%more easy one
%\begin{IEEEeqnarray}{rCl}
%\lefteqn{\sum_v P_V(v)
%D(\pi_{U|V}(\cdot|v)\|Q_{U|V}(\cdot|v))}\nonumber\\
%&\geq&D\left(
%\sum_v P_V(v)\pi_{U|V}(\cdot|v)
%\middle\|
%\sum_v P_V(v)Q_{U|V}(\cdot|v)
%\right)\nonumber\\
%&=&D(P_U\|T_U).
%\IEEEeqnarraynumspace
%\end{IEEEeqnarray}
\end{IEEEproof}

\begin{example}

We illustrate our findings at hand of the following binary example. Consider binary source alphabets
$\mathcal{U}=\mathcal{V}=\{0,1\}$, with joint distributions
\begin{IEEEeqnarray}{rCl}
P_{UV}
&=&
\begin{pmatrix}
0.40 & 0.10\\
0.20 & 0.30
\end{pmatrix},
\\
Q_{UV}
&=&
\begin{pmatrix}
0.15 & 0.25\\
0.35 & 0.25
\end{pmatrix},
\end{IEEEeqnarray}
Let $\Gamma_{Y|X}$ be a binary symmetric
channel (BSC) with crossover probability $p\in(0.4, 0.6)$, and 
$\Gamma_{Z|X}$  a BSC with crossover probability $q=0.38$.

Fig.~\ref{fig:numerical_exponents} depicts the exponents
$\theta_{\textnormal{covert}, \mathcal{H}=0}^{*}(\epsilon)$ (solid blue line) and $\theta_{\textnormal{covert,both}}^{*}(\epsilon)$ (dashed red line) in function of the BSC $\Gamma_{Y|X}$ crossover probability $p$. For comparison, the figure also shows the local exponent at the decision center $D(P_V||Q_V)$ (dotted black line), the Stein-exponent over a partially-connected channel $E_1$ (solid black line), as well as the achievable lower bound on the optimal Stein-exponent without covertness constraints established in \cite[Theorem~1]{Salehkalaibar2020}. 
For this numerical evaluation, the auxiliary
alphabets $\mathcal A$ and $\mathcal T$ (\cite[Theorem~1]{Salehkalaibar2020}) are
restricted to be binary for simplicity.

We observe that the two exponents $\theta_{\textnormal{covert}, \mathcal{H}=0}^{*}(\epsilon)$  and $\theta_{\textnormal{covert,both}}^{*}(\epsilon)$ both improve over the local exponent $D(P_V\|Q_V)$,  unless the BSC($p$) is degenerate $p=1/2$. This observation is consistent with Corollary~\ref{sec:improve_local},  because in this example $T_U \notin \{P_U, Q_U\}$.  The figure further illustrates that in this example the exponent $\theta_{\textnormal{covert}, \mathcal{H}=0}^{*}(\epsilon)$  coincides with the exponent $E_1$ (i.e., the covert Stein-exponent over a partially-connected DMC) when the BSC $\Gamma_{Y|X}$ is sufficiently reliable, and decreases as the channel becomes noisier. Moreover, when covertness is imposed under both hypotheses, the additional constraint reflected by $E_3$ further limits the error exponent; in this case, the partially-connected exponent $E_1$ cannot even be achieved as the BSC cross-over parameter $p\to 0$. Finally, the exponents are also compared to the exponents without a covertness constraint. We observe that they coincide when the covert Stein-exponents are limited by the channel-dependent $E_2$.

\begin{figure}[t]
\centering
\includegraphics[width=\columnwidth]{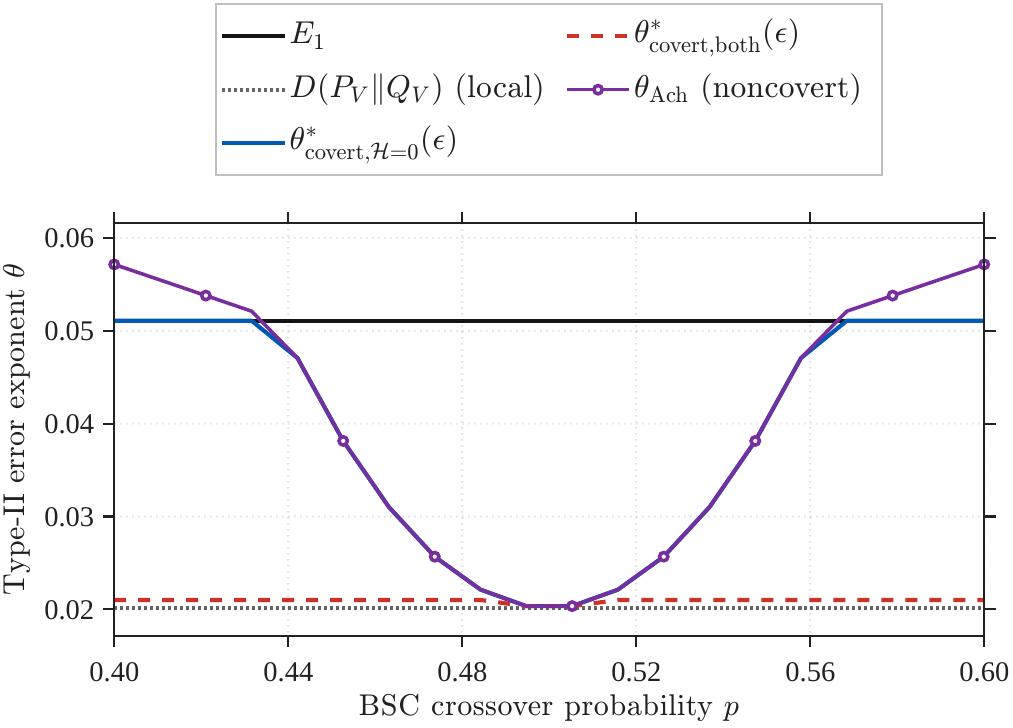}
\caption{Stein-exponents $\theta_{\textnormal{covert}, \mathcal{H}=0}^{*}(\epsilon)$ and $\theta_{\textnormal{covert,both}}^{*}(\epsilon)$ as functions of the crossover
probability $p$, and comparison to the local exponent $D(P_V\|Q_V)$, the Stein-exponent $E_1$ of a partially-connected DMC $\Gamma_{Y|X}$, and the \emph{achievable}  Stein-exponent without covertness constraint of  \cite{Salehkalaibar2020}.}
\label{fig:numerical_exponents}
\end{figure}

    \end{example} 
    
\section{Proof of Achievability Results in Theorem~\ref{sec:theo1} and Theorem~\ref{sec:theo2}} \label{sec:proof_achiev}

\subsection{Achievability of \eqref{eq:Ia}}\label{sec:achiev_part_h0}

\subsubsection{Coding Scheme}
Fix  a small number $\mu >0$ and let $x_1,y^*$ be as in Definition \ref{def:class_chan}.

 Choose further a function $k(\cdot)$ satisfying   
\begin{subequations}\label{eq:sublinear}
\begin{IEEEeqnarray}{rCl}
	\lim_{n\to \infty}k(n)&= &\infty \\
	\lim_{n\to \infty}\frac{k(n)}{n}&= &0. 
\end{IEEEeqnarray}
\end{subequations}We propose a scheme where the sensor uses only the first $k(n)$ channel uses, and sends 0 during the rest of the time.

For ease of notation we will also write $k$ instead of $k(n)$. The scheme works as follows

\noindent\underline{Sensor:} If $U^n \in \mathcal{T}^{(n)}_{\mu}(P_{U})$, the sensor sends $X^k=0^k$. Otherwise, it sends $X^k=x_1^k$.

\noindent\underline{Decision Center:} If at least one of the channel outputs is $y^*$ and if  $V^n \in \mathcal{T}^{(n)}_{\mu}(P_{V})$, then it declares $\hat{\mathcal{H}}=0$. Otherwise, it declares $\hat{\mathcal{H}}=1$. 

\subsubsection{Analysis of error probabilities} As shown in \cite{bounhar2024covert},  the Type-I error probability $\alpha_n$ tends to $0$ as $n \rightarrow \infty$ and $\mu \rightarrow 0$, while the Type-II error exponent exceeds $E_1$.
It remains to verify that the scheme satisfies the covertness  constraint \eqref{eq:cov_h0}. 

\medskip
\subsubsection{Analysis of the covertness constraint}\label{sec:cov_part_h0}
Define 
\begin{equation} 
\rho_{n,0} \triangleq \Pr[U^n \notin \mathcal{T}^{(n)}_{\mu}(P_{U}) |\mathcal{H}=0],
\end{equation}
and notice that 
by   \cite[Remark to Lemma 2.12]{Csiszarbook}: 
\begin{equation} 
\rho_{n,0} \leq 2|\mathcal U| 
 e^{-2n\mu^2}.
\end{equation} 

Then, by the definition of $Z^n$, its law $P_{Z^n|\mathcal{H}=0}$ is given by
\begin{equation}
P_{Z^n|\mathcal{H}=0}=(1-\rho_{n,0})\Gamma_{Z|X}^{\otimes n}(\cdot|0^n)+\rho_{n,0}\Gamma_{Z|X}^{\otimes n}(\cdot|[x_1^k,0^{n-k}]).
\end{equation}

Since  the last $n-k$ remaining channel uses do not contribute to the divergence, we obtain:
\begin{IEEEeqnarray}{rCl}
\lefteqn{D\left( P_{Z^{k}|\mathcal{H}=0} \,\big\|\, 
 \Gamma_{Z|X}^{\otimes k}(\cdot|0^{k})\right)} \nonumber\\
&{=}& D\Big(
(1-\rho_{n,0})\, \Gamma_{Z|X}^{\otimes k}(\,\cdot\,|0^{k}) \label{eq:cov_part_h0_beg}\\
&&  \hspace{20mm} + \ \rho_{n,0}\, \Gamma_{Z|X}^{\otimes k}(\,\cdot\,|x_1^{k})
\,\Big\|\, \Gamma_{Z|X}^{\otimes k}(\,\cdot\,|0^{k})
\Big) \nonumber\\
&\stackrel{(a)}{\le} & \ \rho_{n,0}\, D\Big(
\Gamma_{Z|X}^{\otimes k}(\,\cdot\,|x_1^{k})
\,\Big\|\, \Gamma_{Z|X}^{\otimes k}(\,\cdot\,|0^{k})
\Big) \nonumber\\
&=& k \, \rho_{n,0}\, D\Big(
\Gamma_{Z|X}(\,\cdot\,|x_1)
\,\Big\|\, \Gamma_{Z|X}(\,\cdot\,|0)
\Big) \nonumber\\
&\leq& 2|\mathcal U|  k \,  e^{-2n\mu^2} D\Big(
\Gamma_{Z|X}(\,\cdot\,|x_1)
\,\Big\|\, \Gamma_{Z|X}(\,\cdot\,|0)
\Big)\label{eq:cov_part_h0_end}
\end{IEEEeqnarray}
where $(a)$ follows from the convexity of relative entropy in its first argument.

We can conclude that the right-hand side of \eqref{eq:cov_part_h0_end} tends to $0$ exponentially fast in $n$, thus satisfying the covertness constraint \eqref{eq:cov_h0}.

%Notice that our proposed scheme does not require the usage of the shared secret key $S$ between the transmitter and the receiver. 

%\textcolor{red}{ When the covertness constraint is imposed under both hypotheses, the encoding scheme is modified only slightly: in addition to the above procedure, the sensor also transmits $X^k=0^k$ whenever $U^n \in \mathcal{T}_{\mu}^{(n)}(Q_U)$. However, the optimal Stein-exponent is no longer improved by communication and coincides with the local test,
%\begin{equation}
%\theta^{*}_{\mathrm{Covert}}(\epsilon) = D(P_V \| Q_V).
%<\end{equation} }

\subsection{Achievability of \eqref{eq:Ib}}\label{sec:one_ach_full}

\subsubsection{Coding Scheme}
%Fix a small constant $\mu>0$.% and choose
%\begin{equation}
%x_1\in\arg\max_{x\in\mathcal{X}\setminus\{0\}}D(\Gamma_{Y|X=0}\|\Gamma_{Y|X=x}).
%\end{equation}
Fix a small constant $\mu > 0$.
  
\noindent\underline{Sensor:} If $U^n \in \mathcal{T}^{(n)}_{\mu}(P_{U})$, the sensor sends $X^n=0^n$. Otherwise, it sends $X^n=x_1^n$.

\noindent\underline{\textit{Decision Center:}} 
If 
$V^n \in \mathcal{T}^{(n)}_{\mu}(P_{V})$ 
and 
$Y^n \in \mathcal{T}^{(n)}_{\mu}(\Gamma_{Y|X=0})$,  it declares $\hat{\mathcal{H}} = 0$.  
Otherwise,  $\hat{\mathcal{H}} = 1$.
\medskip

\subsubsection{Analysis of covertness constraint}
The covertness constraint can be analyzed by following the same procedure as in Section \ref{sec:cov_part_h0} . We thus obtain:
\begin{IEEEeqnarray}{rCl}
\lefteqn{D\left( P_{Z^{n}|\mathcal{H}=0} \,\big\|\, 
 \Gamma_{Z|X}^{\otimes n}(\cdot|0^{n})\right)} \label{eq:cov_full_h0_beg}\\
&{=}& D\Big(
(1-\rho_{n,0})\, \Gamma_{Z|X}^{\otimes n}(\,\cdot\,|0^{n}) \\
&&  \hspace{20mm} + \ \rho_{n,0}\, \Gamma_{Z|X}^{\otimes n}(\,\cdot\,|x_1^{n})
\,\Big\|\, \Gamma_{Z|X}^{\otimes n}(\,\cdot\,|0^{n})
\Big) \nonumber\\
&{\le} & \ \rho_{n,0}\, D\Big(
\Gamma_{Z|X}^{\otimes n}(\,\cdot\,|x_1^{n})
\,\Big\|\, \Gamma_{Z|X}^{\otimes n}(\,\cdot\,|0^{n})
\Big) \nonumber\\
&=& n \, \rho_{n,0}\, D\Big(
\Gamma_{Z|X}(\,\cdot\,|x_1)
\,\Big\|\, \Gamma_{Z|X}(\,\cdot\,|0)
\Big) \nonumber\\
&\leq& 2|\mathcal U|  n \,  e^{-2n\mu^2} D\Big(
\Gamma_{Z|X}(\,\cdot\,|x_1)
\,\Big\|\, \Gamma_{Z|X}(\,\cdot\,|0)
\Big)\label{eq:cov_full_h0_end},
\end{IEEEeqnarray}
satisfying the desired covertness constraint under $\mathcal{H}=0$ as $n$ tends to infinity.

\subsubsection{Analysis of error probabilities} \label{sec:error_proba_full_h0}
The analysis of the Type-I and Type-II error probabilities follows exactly the same steps as in \cite{bounhar2026distributed}. Nevertheless, for completeness, we reproduce the analysis here.

\noindent\underline{Analysis of $\alpha_n$:} We have: 
\begin{IEEEeqnarray}{rCl}
\lefteqn{1-\alpha_n}\nonumber\\
&=&\Pr\left[\hat{\mathcal{H}}=0 |\mathcal{H}=0\right]\label{eq:01}\\
&=&\Pr\left[ Y^n \in  \mathcal{T}_{\mu}^{(n)}(\Gamma_{Y|X}(\cdot|0)) \; \textnormal{and} \; V^n \in \mathcal{T}_{\mu}^{(n)}(P_{V}) |\mathcal{H}=0\right] \nonumber\\
 \\
&\geq & \Pr\big[ Y^n \in  \mathcal{T}_{\mu}^{(n)}(\Gamma_{Y|X}(\cdot|0)) \; \textnormal{and} \; \nonumber \\
&& \hspace{1.4cm} U^n \in  \mathcal{T}_{\mu_n}^{(n)}(P_{U})\; \textnormal{and} \; V^n \in \mathcal{T}_{\mu}^{(n)}(P_{V}) |\mathcal{H}=0\big]\nonumber\\\\ 
& =  & \Pr[ Y^n \in  \mathcal{T}_{\mu}^{(n)}(\Gamma_{Y|X}(\cdot|0)) |X^n=0^n]  \nonumber \\
&&  \cdot\Pr\left[ U^n \in  \mathcal{T}_{\mu}^{(n)}(P_{U})\; \textnormal{and} \; V^n \in \mathcal{T}_{\mu}^{(n)}(P_{V}) |\mathcal{H}=0\right] .\label{eq:fin}
\end{IEEEeqnarray}
Notice that by the weak law of large numbers, both probabilities in \eqref{eq:fin} tend to 1 and thus
\begin{IEEEeqnarray}{rCl}
\lim_{n\to \infty} \alpha_n=0.
\end{IEEEeqnarray}

\medskip
\noindent\underline{Analysis of $\beta_n$ and $\theta$:}
Similarly to above, we have:
\begin{IEEEeqnarray}{rCl}
\beta_n&=&\Pr\left[\hat{\mathcal{H}}=0 |\mathcal{H}=1\right]\\
& = &  \Pr\Big[ Y^n \in  \mathcal{T}_{\mu}^{(n)}(\Gamma_{Y|X}(\cdot|0)) \; \textnormal{and} \; \nonumber \\
&& \hspace{1.2cm} U^n \in  \mathcal{T}_{\mu}^{(n)}(P_{U})\; \textnormal{and} \; V^n \in \mathcal{T}_{\mu}^{(n)}(P_{V}) \; \big|\;\mathcal{H}=1\Big] \nonumber \\
&& +  \Pr\Big[ Y^n \in  \mathcal{T}_{\mu}^{(n)}(\Gamma_{Y|X}(\cdot|0)) \; \textnormal{and} \;\nonumber \\
&& \hspace{1.2cm} U^n \notin  \mathcal{T}_{\mu}^{(n)}(P_{U})\; \textnormal{and} \; V^n \in \mathcal{T}_{\mu}^{(n)}(P_{V}) \; \big|\;\mathcal{H}=1\Big] \nonumber \\\\
& \leq  & \Pr\left[ U^n \in  \mathcal{T}_{\mu}^{(n)}(P_{U})\; \textnormal{and} \; V^n \in \mathcal{T}_{\mu}^{(n)}(P_{V}) \; \big|\;\mathcal{H}=1\right] \nonumber\\
&& +  \Pr\left[ Y^n \in  \mathcal{T}_{\mu}^{(n)}(\Gamma_{Y|X}(\cdot|0))\; \big|\;X^n=x_1^n\right]  \nonumber \\
&&  \cdot\Pr\left[  V^n \in \mathcal{T}_{\mu}^{(n)}(P_{V}) |\mathcal{H}=1\right] \\
&\leq &     (n+1)^{|\mathcal{U}||\mathcal{V}|} e^{-n \min D( \pi_{UV}\|Q_{UV})} \nonumber\\ 
&&+  (n+1)^{|\mathcal{Y}|} e^{-n \min D( \pi_{Y}\|\Gamma_{Y|X=x_1})}  \nonumber \\
&& \cdot    (n+1)^{|\mathcal{V}|} e^{-n \min D( \pi_{V}\|Q_{V})} ,
\end{IEEEeqnarray}
where the first  minimum is over all pmfs $\pi_{UV}$ with marginals  that are $\mu$-close to $P_U$ and $P_V$, and the latter two minima are over all pmfs $\pi_Y$ and $\pi_V$ that are $\mu$-close to $\Gamma_{Y|X=0}$ and $P_V$, respectively. Letting $n\to \infty$ and picking $\mu>0$ sufficiently small, by standard arguments we have: 
\begin{IEEEeqnarray}{rCl}
\lim_{n\to \infty}  \min D( \pi_{Y}\|\Gamma_{Y|X=x_1})& =& D(\Gamma_{Y|X=0} \| \Gamma_{Y|X=x_1}) \IEEEeqnarraynumspace\\
\lim_{n\to \infty}  \min D( \pi_{V}\|Q_{V})&= & D(P_V\| Q_V) \\
\lim_{n\to \infty}  \min D( \pi_{UV}\|Q_{UV})&= & E_1.  
\end{IEEEeqnarray}
We can thus conclude that for the proposed scheme: 
\begin{IEEEeqnarray}{rCl}
\lim_{n\to \infty} -\frac{1}{n} \log \beta_n \geq   \min\{E_1, E_2\}. 
\end{IEEEeqnarray}

This establishes achievability of the exponent in \eqref{eq:Ib}.

\subsection{Achievability of \eqref{eq:bothpart}}\label{sec:dual_covertness_partial}

\subsubsection{Coding scheme}

Let  $x_1,y^*$ be as in Definition \ref{def:class_chan}.
%\begin{equation}
%\Gamma_{Y|X}(y^*|0)>0,\qquad \Gamma_{Y|X}(y^*|x_1)=0.
%\end{equation}
Choose a sequence of positive integers $\{m_n\}_{n\geq1}$ such that
\begin{equation}
m_n\to\infty,\qquad \frac{m_n^2}{n}\to0.
\end{equation}

Let $\mathcal{K}\triangleq\{s\subseteq\{1,\ldots,n\}:|s|=m_n\}$. In this case, $|\mathcal{K}|= \binom{n}{m_n}$ and  the   key size $\log_2 |\mathcal{K}|= \log_2 \binom{n}{m_n}$ is in the order of $O(m_n \log (n))$ bits. 
\iffalse
prrof:
By the definition of the binomial coefficient,
\begin{IEEEeqnarray}{rCl}
\binom{n}{m_n}
&=&\frac{n!}{m_n!(n-m_n)!}\nonumber\\
&=&\frac{n(n-1)\cdots(n-m_n+1)}{m_n!}\nonumber\\
&\leq&\frac{n^{m_n}}{m_n!}\nonumber\\
&\leq&n^{m_n}.
\end{IEEEeqnarray}
Taking the logarithm to base $2$ on both sides yields
\begin{IEEEeqnarray}{rCl}
\log_2\binom{n}{m_n}
&\leq&\log_2(n^{m_n})\nonumber\\
&=&m_n\log_2 n.
\end{IEEEeqnarray}
Consequently,
\begin{IEEEeqnarray}{rCl}
\log_2|\mathcal K|&=&O(m_n\log n).
\end{IEEEeqnarray}
\fi
%The sensor and the detector share a secret key $S$, independent of $(U^n,V^n)$, which is uniformly distributed over $\mathcal{K}$, i.e.,
%\begin{equation}
%P_S(s)=\frac{1}{\binom{n}{m_n}},\qquad s\in\mathcal{K}.
%\end{equation}

For each $s\in\mathcal{K}$, define $x^n(s)\in\mathcal{X}^n$ as
\begin{equation}
x_i(s)=\begin{cases}x_1, & i\in s,\\ 0, & i\notin s.\end{cases}
\end{equation}

\noindent\underline{Sensor:}
Given $(U^n,S)=(u^n,s)$, the sensor transmits
\begin{equation}
f^{(n)}(u^n,s)=\begin{cases}0^n, & u^n\in\mathcal{T}_{\mu}^{(n)}(P_U),\\ x^n(s), & u^n\notin\mathcal{T}_{\mu}^{(n)}(P_U).\end{cases}
\end{equation}

\noindent\underline{Decision Center:}
Upon observing $(V^n,Y^n,S)=(v^n,y^n,s)$, the decision center decides $\mathcal{\hat{H}}=0$ if
\begin{equation}
v^n\in\mathcal{T}_{\mu}^{(n)}(P_V)\quad\text{and}\quad \exists\,i\in s:\ y_i=y^*,
\end{equation}
and decides $\mathcal{\hat{H}}=1$ otherwise.

\subsubsection{Analysis of the covertness constraint}

%The following lemma will be useful to establish the covertness constraint.
%\begin{lemma}\label{lem:div_covert_bound}
%Consider the coding scheme described above, where the secret key $S$ is uniformly distributed over $\mathcal{K}$. For $H\in\{0,1\}$, 
Let $\rho_{n,H}\triangleq\Pr[U^n\notin\mathcal{T}_{\mu}^{(n)}(P_U)\mid\mathcal{H}=H]$. As shown in Appendix~\ref{app:proof_lemma_div},
%\begin{IEEEeqnarray}{rCl}
%\lefteqn{D(P_{Z^n|\mathcal{H}=H}\|\Gamma_{Z|X=0}^{\otimes n})}\nonumber\\
%&\leq&\rho_{n,H}^2\left[\left(1+\frac{m_n}%{n}\chi^2(\Gamma_{Z|X=x_1}\|\Gamma_{Z|X=0})\right)^{m_n}-1\right],
%\IEEEeqnarraynumspace\label{eq:both_covertness_bound}
%\end{IEEEeqnarray}
%\end{lemma}
%
%\begin{IEEEproof}
%See Appendix~\ref{app:proof_lemma_div}.
%\end{IEEEproof}
%
%By  Lemma~\ref{lem:div_covert_bound}, 
for each $H\in\{0,1\}$:
\begin{IEEEeqnarray}{rCl}
\lefteqn{D(P_{Z^n|\mathcal{H}=H}\|\Gamma_{Z|X=0}^{\otimes n})}\nonumber\\
&\leq&\rho_{n,H}^2\left[\left(1+\frac{m_n}{n}\chi^2(\Gamma_{Z|X=x_1}\|\Gamma_{Z|X=0})\right)^{m_n}-1\right] \label{eq:TT}\\
&\leq&\exp\left(\frac{m_n^2}{n}\chi^2(\Gamma_{Z|X=x_1}\|\Gamma_{Z|X=0})\right)-1,\label{eq:both_covertness_final}
\end{IEEEeqnarray}
where the last inequality holds because $\rho_{n,H}\leq1$ and $1+t\leq e^t$.

Since $m_n^2/n\to0$, the right-hand side of \eqref{eq:both_covertness_final} tends to zero. Consequently, for both $H=0$ and $H=1$, we have
\begin{IEEEeqnarray}{rCl}
\lim_{n\to\infty}D\left(P_{Z^n|\mathcal{H}=H}\|\Gamma_{Z|X=0}^{\otimes n}\right)=0,
\end{IEEEeqnarray}
and the proposed coding scheme satisfies the covertness constraints under both hypotheses.

\subsubsection{Analysis of error probabilities}

Define $\gamma_{0} \triangleq \Gamma_{Y|X}(y^*|0)$ and $\gamma_{x_1} \triangleq \Gamma_{Y|X}(y^*|x_1)$.

\noindent\underline{Analysis of $\alpha_n$:}
\begin{IEEEeqnarray}{rCl}
\lefteqn{1-\alpha_n} \nonumber\\
&=& \Pr\left[\hat{\mathcal{H}}=0 \mid \mathcal{H}=0\right]\nonumber \\
&=& \Pr\left[\exists i\in S:Y_i=y^* \;\textnormal{and}\; V^n\in\mathcal{T}^{(n)}_{\mu}(P_V)\mid\mathcal{H}=0\right] \nonumber\\
&\geq& \Pr\big[ \exists i\in S:Y_i=y^*,\; V^n\in\mathcal{T}^{(n)}_{\mu}(P_V), \nonumber\\
&&\hspace{1cm} U^n\in\mathcal{T}^{(n)}_{\mu}(P_U)\mid\mathcal{H}=0\big] \nonumber \\
&=& \Pr\left[V^n\in\mathcal{T}^{(n)}_{\mu}(P_V),\; U^n\in\mathcal{T}^{(n)}_{\mu}(P_U)\mid\mathcal{H}=0\right]\nonumber\\
&&\cdot\Pr\left[\exists i\in S:Y_i=y^*\mid U^n\in\mathcal{T}^{(n)}_{\mu}(P_U),\mathcal{H}=0\right] \nonumber\\
&=& \Pr\left[V^n\in\mathcal{T}^{(n)}_{\mu}(P_V),\; U^n\in\mathcal{T}^{(n)}_{\mu}(P_U)\mid\mathcal{H}=0\right]\nonumber\\
&&\cdot\left(1-(1-\gamma_0)^{m_n}\right).\label{eq:alpha_both_key}
\end{IEEEeqnarray}

Since $\gamma_0\in(0,1]$ and $m_n\to\infty$, we have
\begin{IEEEeqnarray}{rCl}
\lim_{n\to\infty}(1-\gamma_0)^{m_n}=0.
\end{IEEEeqnarray}
Moreover, by the weak law of large numbers,
\begin{IEEEeqnarray}{rCl}
\lim_{n\to\infty}\Pr\left[U^n\in\mathcal{T}^{(n)}_{\mu}(P_U),\;V^n\in\mathcal{T}^{(n)}_{\mu}(P_V)\mid\mathcal{H}=0\right]=1. \nonumber \\
\end{IEEEeqnarray}
Combining these limits with \eqref{eq:alpha_both_key}, we conclude that the type-I error probability vanishes:
\begin{IEEEeqnarray}{rCl}
\lim_{n\to\infty}\alpha_n=0.
\end{IEEEeqnarray}

\noindent\underline{Analysis of $\beta_n$ and $\theta$:}
We have:
\begin{IEEEeqnarray}{rCl}
\beta_n
&=&\Pr\left[\hat{\mathcal{H}}=0\mid\mathcal{H}=1\right] \nonumber\\
&=&\Pr\big[\exists i\in S:Y_i=y^*\;\textnormal{and}\;V^n\in\mathcal{T}^{(n)}_{\mu}(P_V)\mid\mathcal{H}=1\big]\nonumber\\
&=&\Pr\big[\exists i\in S:Y_i=y^*\;\textnormal{and}\;V^n\in\mathcal{T}^{(n)}_{\mu}(P_V)\nonumber\\
&&\hspace{1cm}\textnormal{and}\;X^n=0^n\;\textnormal{and}\;U^n\in\mathcal{T}^{(n)}_{\mu}(P_U)\mid\mathcal{H}=1\big]\\
&+&\Pr\big[\exists i\in S:Y_i=y^*\;\textnormal{and}\;V^n\in\mathcal{T}^{(n)}_{\mu}(P_V)\nonumber\\
&&\hspace{1cm}\textnormal{and}\;X^n=x^n(S)\;\textnormal{and}\;U^n\notin\mathcal{T}^{(n)}_{\mu}(P_U)\mid\mathcal{H}=1\big]\nonumber\\
&=&\Pr\big[V^n\in\mathcal{T}^{(n)}_{\mu}(P_V)\;\textnormal{and}\;U^n\in\mathcal{T}^{(n)}_{\mu}(P_U)\mid\mathcal{H}=1\big]\nonumber\\
&&\cdot\underbrace{\Pr\big[\exists i\in S:Y_i=y^*\mid X^n=0^n\big]}_{=1-(1-\gamma_0)^{m_n}}\nonumber\\
&+&\Pr\big[V^n\in\mathcal{T}^{(n)}_{\mu}(P_V)\;\textnormal{and}\;U^n\notin\mathcal{T}^{(n)}_{\mu}(P_U)\mid\mathcal{H}=1\big]\nonumber\\
&&\cdot\underbrace{\Pr\big[\exists i\in S:Y_i=y^*\mid X^n=x^n(S)\big]}_{=1-(1-\gamma_{x_1})^{m_n}=0}\\
&=&\Pr\big[V^n\in\mathcal{T}^{(n)}_{\mu}(P_V)\;\textnormal{and}\;U^n\in\mathcal{T}^{(n)}_{\mu}(P_U)\mid\mathcal{H}=1\big]\nonumber\\
&&\cdot\left(1-(1-\gamma_0)^{m_n}\right)\nonumber\\
&\leq&\Pr\big[V^n\in\mathcal{T}^{(n)}_{\mu}(P_V)\;\textnormal{and}\;U^n\in\mathcal{T}^{(n)}_{\mu}(P_U)\mid\mathcal{H}=1\big]\nonumber\\
&\leq&(n+1)^{|\mathcal{U}||\mathcal{V}|}e^{-n\min D(\pi_{UV}\|Q_{UV})},\label{eq:beta_both_key}
\end{IEEEeqnarray}
where the  minimum is over all pmfs $\pi_{UV}$ with marginals  that are $\mu$-close to $P_U$ and $P_V$. Letting $n\to\infty$ and  $\mu\downarrow0$, standard arguments based on the method of types yield
\begin{IEEEeqnarray}{rCl}
\varliminf_{n\to\infty}-\frac{1}{n}\log\beta_n\geq E_1.
\end{IEEEeqnarray}
\smallskip

\subsection{Achievability of \eqref{eq:bothfull}}
\label{rem:dual_covertness_fully}
\subsubsection{Coding scheme}
Apply the same coding scheme as in Subsection~\ref{sec:one_ach_full}, except that the sensor also sends $X^n=0^n$ when $U^n \in \mathcal{T}_{\mu}^{(n)}(Q_{U})$. Thus, the sensor transmits $X^n=0^n$ whenever $U^n\in\mathcal{T}_{\mu}^{(n)}(P_U)\cup\mathcal{T}_{\mu}^{(n)}(Q_U)$, and $X^n=x_1^n$ otherwise.

\subsubsection{Analysis of the covertness constraint}
Analysis of the covertness constraint under $\mathcal{H}=0$ and $\mathcal{H}=1$  follows the 
Steps in \eqref{eq:cov_full_h0_beg}--\eqref{eq:cov_full_h0_end}. Thus, the covertness constraint is satisfied under both hypotheses.

\subsubsection{Analysis of error probabilities} 
\ \  \\
\noindent\underline{Analysis of $\alpha_n$:}
One can follow exactly the same analysis steps \eqref{eq:01}--\eqref{eq:fin}  as in Section \ref{sec:error_proba_full_h0}.

\noindent\underline{Analysis of $\beta_n$:}
Again, the analysis  follows exactly the same steps
as in \cite{bounhar2026distributed}. Nevertheless, for completeness, we reproduce the analysis here. We have:
\begin{IEEEeqnarray}{rCl}
\beta_n
&=&\Pr\left[\hat{\mathcal{H}}=0 |\mathcal{H}=1\right]\\
& \leq &  \Pr\big[ Y^n \in  \mathcal{T}_{\mu}^{(n)}(\Gamma_{Y|X}(\cdot|0)) \; \textnormal{and} \;  V^n \in \mathcal{T}_{\mu}(P_{V})\nonumber \\
&& \hspace{3cm}  \textnormal{and} \; U^n \in  \mathcal{T}_{\mu}^{(n)}(P_{U}) |\mathcal{H}=1\big] \nonumber \\
&  &+   \Pr\big[ Y^n \in  \mathcal{T}_{\mu}^{(n)}(\Gamma_{Y|X}(\cdot|0))\; \textnormal{and} \; V^n \in \mathcal{T}_{\mu}(P_{V})  \nonumber \\
&& \hspace{3cm} \; \textnormal{and} \;U^n \in  \mathcal{T}_{\mu}^{(n)}(Q_{U}) |\mathcal{H}=1\big] \nonumber \\
&& +  \Pr\Big[ Y^n \in  \mathcal{T}_{\mu_n}^{(n)}(\Gamma_{Y|X}(\cdot|0)) \; \textnormal{and} \; V^n \in \mathcal{T}_{\mu}^{(n)}(P_{V})\nonumber \\
&& \hspace{1.2cm}\; \textnormal{and} \; U^n \notin ( \mathcal{T}_{\mu}^{(n)}(P_{U}) \cup \mathcal{T}_{\mu}^{(n)}(Q_{U}) )  |\mathcal{H}=1\Big] \nonumber\\
& \leq  & \Pr\left[ U^n \in  \mathcal{T}_{\mu}^{(n)}(P_{U})\; \textnormal{and} \; V^n \in \mathcal{T}_{\mu}^{(n)}(P_{V}) |\mathcal{H}=1\right] \nonumber\\
&&+  \Pr\left[ U^n \in  \mathcal{T}_{\mu}^{(n)}(Q_{U})\; \textnormal{and} \; V^n \in \mathcal{T}_{\mu}^{(n)}(P_{V}) |\mathcal{H}=1\right] \nonumber\\
&& +  \Pr[ Y^n \in  \mathcal{T}_{\mu}^{(n)}(\Gamma_{Y|X}(\cdot|0)) |X^n=x_1^n]  \nonumber \\
&&  \cdot\Pr\left[  V^n \in \mathcal{T}_{\mu}^{(n)}(P_{V}) |\mathcal{H}=1\right] \\
&\leq &   (n+1)^{|\mathcal{U}||\mathcal{V}|} e^{-n \min D( \pi_{UV}\|Q_{UV})}  \nonumber\\
& & + (n+1)^{|\mathcal{U}||\mathcal{V}|} e^{-n \min D( \pi_{UV}\|Q_{UV})}  \nonumber\\
& &+e^{-n \min D( \pi_{Y}\|\Gamma_{Y|X=x_1})}     (n+1)^{|\mathcal{V}|} e^{-n \min D( \pi_{V}\|Q_{V})} ,\nonumber\\
\end{IEEEeqnarray}
where  the first two minimima are  over all pmfs $\pi_{UV}$ with marginals  that are $\mu$-close to $P_U$ and $P_V$ and to $Q_U$ and $P_V$, respectively,  and the latter two minima are over all pmfs $\pi_Y$ and $\pi_V$ that are $\mu$-close to $\Gamma_{Y|X=0}$ and $P_V$, respectively.  Letting $n\to \infty$, by standard arguments we have: 
\begin{IEEEeqnarray}{rCl}
\varliminf_{n\to \infty} -\frac{1}{n} \log \beta_n \geq  \min\{E_1, E_2, E_3\}. 
\end{IEEEeqnarray}
\smallskip

\section{Proof of Converse Results in Theorem~\ref{sec:theo1} and Theorem~\ref{sec:theo2}} \label{sec:proof_conv}

\subsection{Common Definitions and Lemmas}
Before starting the converse proofs, we introduce some
common definitions and establish some preliminary results
for given encoding and decision functions
$(f^{(n)},g^{(n)})$.

Define for each $n$ and each hypothesis $H\in\{0,1\}$ :
 \begin{IEEEeqnarray}{rCl}
\label{eq:def_alpha_n_i_0}
a_{n,i}^H&\triangleq &\Pr[ X_{i} \neq 0 |\mathcal{H}= H],  \quad  i\in\{1,\ldots, n\},%\\
%b_{n,i}^H(x) &\triangleq &\Pr[ X_{i} = x |X_{i} \neq 0, \mathcal{H}= H], \quad x \in \mathcal{X}\backslash \{0\},\IEEEeqnarraynumspace
\end{IEEEeqnarray}
and 
\begin{equation}\label{eq:eps_n}
\bar a_n^{H} \triangleq % \max \left \{ 
\sqrt{\frac{1}{n} \sum_{i=1}^n a_{n,i}^{H} }.% \frac{1}{\log n} \right\} , 
\end{equation}
%so that $\bar a_n$ vanishes but more slowly  than $\frac{1}{n} \sum_{i=1}^n a_{n,i}^{0}$. 
Define  further the set of all low-weight inputs
\begin{IEEEeqnarray}{rCl}
\tilde{\mathcal{X}}^n_0 &\triangleq &\big\{ x^n \in \mathcal{X}^n \colon w_{\textnormal{H}}(x^n) \leq \bar a_n^{0} \cdot n\big\}, \label{eq:TX}\\
\tilde{\mathcal{X}}^n_{\textnormal{both}}& \triangleq& \big\{ x^n \in \mathcal{X}^n \colon w_{\textnormal{H}}(x^n) \leq \max\{ \bar a_n^{0} , \bar a_n^{1}\} \cdot n\big\},\label{eq:TXboth}
\end{IEEEeqnarray} 
as well as the corresponding sets of source-key pairs that are encoded into low-weight inputs
\begin{IEEEeqnarray}{rCl}
\widetilde{\mathcal{US}}_0
&\triangleq&
\bigl\{(u^n,s)\in\mathcal U^n \times \mathcal{S}\colon 
f^{(n)}(u^n,s) \in\tilde{\mathcal{X}}^n_0 \bigr\} \label{eq:US}\\
 \widetilde{\mathcal{US}}_{\textnormal{both}}
&\triangleq&
\bigl\{(u^n,s)\in\mathcal U^n \times \mathcal{S}\colon 
 f^{(n)}(u^n,s) \in \tilde{\mathcal{X}}^n_{\textnormal{both}} \bigr\}. \label{eq:USboth}
\end{IEEEeqnarray} 

\begin{lemma}[Covertness Constraint Implies Low-Weight Inputs]\label{lem:lemma_low_weight} 
If covertness is imposed under $\mathcal{H}=0$:
\begin{IEEEeqnarray}{rCl}
\lim_{n\to\infty}\bar  a_n^0 &=& 0, \label{eq:van_a0}\\
\lim_{n\to \infty}  \Pr\left[X^n \notin \tilde{\mathcal{X}}_0^n\, \big| \,\mathcal{H}=0\right]  &=& 0 , \label{eq:high_prob}\\
\lim_{n\to \infty} \frac{1}{n} \log \left| \tilde{\mathcal{X}}_0^n \right| &= & 0.\label{eq:sizeX}
\end{IEEEeqnarray}
Similarly, if the covertness constraint is imposed under both hypotheses, then for ${H}\in\{0,1\}$:
\begin{IEEEeqnarray}{rCl}
\lim_{n\to\infty} \bar a_n^{{H}} &=& 0, \label{eq:van_aboth}\\
\lim_{n\to \infty}  \Pr\left[X^n \notin \tilde{\mathcal{X}}_{\textnormal{both}}^n\, \big| \,\mathcal{H}= H\right]  &=& 0 , \label{eq:high_prob_both}\\
\lim_{n\to \infty} \frac{1}{n} \log \left| \tilde{\mathcal{X}}_{\textnormal{both}}^n \right| &= & 0.\label{eq:sizeX_both}
\end{IEEEeqnarray}
\end{lemma} 
\begin{IEEEproof} 
See Appendix~\ref{app:proof_low_weight}.
\end{IEEEproof} 
\medskip
For each pair $(v^n, s)$, define the acceptance region
\begin{equation}
\mathcal A_n(v^n,s)
\triangleq
\left\{
y^n\in\mathcal Y^n:
g^{(n)}(v^n,y^n,s)=0
\right\},
\end{equation}
and its $r$-blown-up region
\begin{IEEEeqnarray}{rCl}
\lefteqn{\widehat{\mathcal A}_n^{r}(v^n,s)}\nonumber\\
&\triangleq&
\bigl\{\widetilde y^n:\exists y^n\in\mathcal A_n(v^n,s)
\text{ s.t. }d_{\mathrm H}(\widetilde y^n,y^n)\le r\bigr\}.
\end{IEEEeqnarray}
Further, let $B_n(r)$ denote the cardinality of the
$n$-dimensional $r$-ball
\begin{equation}
\mathbb{B}^r_n(y^n):=\{\tilde{y}^n\in \mathcal{Y}^n
\colon d_H(y^n,\tilde{y}^n)\leq r\}.
\label{eq:hammmin_bamm}
\end{equation}
By the standard Hamming-ball cardinality bound
(see, e.g., \cite[Proof of Lemma~5.1]{Csiszarbook}),
\begin{IEEEeqnarray}{rCl}
B_n(r)&\leq&e^{nH_{\textnormal b}(r/n)}|\mathcal Y|^r.
\label{eq:hamming_ball_bound}
\end{IEEEeqnarray}

For fully-connected DMCs, define
\begin{IEEEeqnarray}{rCl}
\gamma_{\min}&\triangleq&
\min_{x\in\mathcal X,y\in\mathcal Y}
\Gamma_{Y|X}(y|x)>0.
\label{eq:E3_gamma_min}
\end{IEEEeqnarray}
Then, for any triplet $(u^n,v^n,s)$, any hypothesis
$H\in\{0,1\}$, we have
\begin{IEEEeqnarray}{rCl}
\lefteqn{\Pr\left[Y^n\in\widehat{\mathcal A}_n^{r}(v^n,s)
\mid U^n=u^n,V^n=v^n,S=s,\mathcal{H}=H\right]}\nonumber\\
&\stackrel{(a)}{\leq}&
\Pr\left[Y^n\in\mathcal A_n(v^n,s)
\mid U^n=u^n,V^n=v^n,S=s,\mathcal{H}=H\right]\nonumber\\
&&\hspace{.2cm}\cdot\gamma_{\min}^{-r}B_n(r)
\label{eq:blowup_counting}\\
&\stackrel{(b)}{\leq}&
\Pr\left[Y^n\in\mathcal A_n(v^n,s)
\mid U^n=u^n,V^n=v^n,S=s,\mathcal{H}=H\right]\nonumber\\
&&\hspace{.2cm}\cdot
e^{nH_{\textnormal b}(r/n)}|\mathcal Y|^r
\gamma_{\min}^{-r},
\label{eq:blowup_probability_bound}
\end{IEEEeqnarray}
where Inequality $(a)$ holds by simple counting arguments
and Inequality $(b)$ follows from
\eqref{eq:hamming_ball_bound}.

\subsection{Upper Bound $E_1$}

The converse is inspired by \cite{bounhar2024covert}. We also assume a stronger setup where the decision center  directly observes the DMC inputs $X^n$ instead of the outputs $Y^n$. A converse result for this new setup implies also a converse for the original setup. Thus, we assume in the following that  \eqref{eq:guess}  can  be replaced by
\begin{equation}\label{eq:guess2}
\hat{\mathcal{H}} =	\tilde g^{(n)}(V^n,X^n, S) \in\{0,1\}.
\end{equation}

Fix any sequence of encoding and modified decision functions $\{f^{(n)},\tilde{g}^{(n)}\}$ satisfying the Type-I error constraint \eqref{eq:P12} and  the covertness constraint \eqref{eq:cov} under $\mathcal{H}=0$.

Then consider:
\begin{IEEEeqnarray}{rCl}
\lefteqn{
 \Pr\left[ \hat{\mathcal{H}}=0, X^n\in \tilde{ \mathcal{X}}_0^n| \mathcal{H}=0\right] }\nonumber \\
& = & \Pr\left[ \hat{\mathcal{H}}=0 | \mathcal{H}=0\right]  -   \Pr\left[ \hat{\mathcal{H}}=0, X^n\notin \tilde{ \mathcal{X}}_0^n| \mathcal{H}=0\right]  \IEEEeqnarraynumspace\\
& \geq & 1- \epsilon -   \Pr[X^n \notin \tilde{\mathcal{X}}_0^n | \mathcal{H}=0],
\end{IEEEeqnarray} 
%where recall that $\tilde{\mathcal{X}}_0$ is defined in \eqref{eq:TX}.

 By Lemma~\ref{lem:lemma_low_weight} and because the chosen sequence of encoding functions satisfies the covertness constraint \eqref{eq:cov} under $H=0$, for any  $\eta \in (0,1-\epsilon)$ and sufficiently large $n$:
\begin{equation}
 \Pr\left[ \hat{\mathcal{H}}=0, X^n\in \tilde{ \mathcal{X}}_0^n \big| \mathcal{H}=0\right] \geq 1- \epsilon-\eta. 
\end{equation}
In particular, there must  be a special sequence $\bar{x}^n\in \tilde{\mathcal{X}}_0^n$   so that: 
\begin{equation}\label{eq:bound}
\Pr\left [ \hat{\mathcal{H}}=0, X^n=\bar{x}^n\Big | \mathcal{H}=0\right] \geq \frac{1- \epsilon-\eta}{|\tilde{\mathcal{X}}_0^n|}. 
\end{equation}

Fix this sequence $\bar{x}^n$ and choose an arbitrary constant $c>1$. Define then the subset $\mathcal{\bar{S}}_{\bar{x}^n} \subseteq \mathcal{K}$
\begin{IEEEeqnarray}{rCl}
\label{eq:set_S_bar}
\mathcal{\bar{S}}_{\bar{x}^n}
&&\triangleq \bigg\{{s}\in \mathcal{K} \colon\nonumber \\
&& \hspace{.7cm}\Pr\left [ \hat{\mathcal{H}}=0, X^n=\bar{x}^n\Big | \mathcal{H}=0 , S={s}\right] \geq 
\frac{1- \epsilon-\eta}{c|\tilde{\mathcal{X}}_0^n|}\bigg\}.\nonumber\\
\end{IEEEeqnarray}

Notice that by Lemma \ref{lem:lemma_low_weight} the cardinality of $|\tilde{\mathcal{X}}_0^n|$  grows sub-exponentially in $n$ and thus the ratio  $\frac{1- \epsilon-\eta}{c|\tilde{\mathcal{X}}_0^n|}$ is of the form $e^{-n \phi_n}$ for a vanishing sequence $\phi_n$. i.e., for any ${s}\in \mathcal{{S}}_{\bar{x}^n}$:
\begin{equation} 
\Pr\left [ \hat{\mathcal{H}}=0, X^n=\bar{x}^n\Big | \mathcal{H}=0 , S={s}\right] \geq e^{-n \phi_n},\label{eq:exp}
\end{equation}
for 
\begin{equation} 
\phi_n \triangleq \frac{1}{n} \ln \frac{c|\tilde{\mathcal{X}}^n_0|}{1- \epsilon-\eta}.
\end{equation} 

We next notice the following lemma: 
\begin{lemma} \label{lem:lemma_ps}
%Consider any sequence of encoding and improved decision functions $\{ f^{(n)},\tilde{g}^{(n)}\}$. 
Consider %further 
a vanishing sequence  $\phi_n \to 0$, as $n\to \infty$. If for each $n$ there exists an input sequence  $\bar{x}^n\in \tilde{\mathcal{X}}_0^n$ and key $\bar{s} \in \mathcal{K}$ satisfying
\begin{equation}
\Pr\left [ \hat{\mathcal{H}}=0, X^n=\bar{x}^n\Big | \mathcal{H}=0 , S={s}\right] \geq e^{-n \phi_n},
\end{equation} 
then 
\begin{IEEEeqnarray}{rCl}    \lefteqn{\hspace{-0.8cm}
\Pr[\hat{\mathcal{H}}=0 , X^n=\bar{x}^n| \mathcal{H}=1, S={s}]} \nonumber\\ 
&\geq
&e^{-n( \min_{\pi_{UV}}D(\pi_{UV}\| Q_{UV} )+\eta_n)},\hspace{1cm}
\end{IEEEeqnarray}
where the minimum is over all pmfs $\pi_{UV}$ that have marginals $P_U$ and $P_V$  and $\eta_n$ is a vanishing sequence in $n$ that does not depend on $\bar{x}^n$ nor $\bar{s}$ but only on $P_{UV},Q_{UV}$ and the sequence $\phi_n$.
\end{lemma} 
\begin{IEEEproof} 
See Appendix~\ref{app:proof_lemma_ps}.
\end{IEEEproof}

Combining \eqref{eq:exp} with Lemma~\ref{lem:lemma_ps} implies that there exists a vanishing sequence $\eta_n$ so that for any $\bar{s}\in \mathcal{\bar{S}}_{\bar{x}^n}$:
\begin{IEEEeqnarray}{rCl}
\lefteqn{\Pr[\hat{\mathcal{H}}=0 , X^n=\bar{x}^n| \mathcal{H}=1, S=\bar{s}] } \nonumber \\
& \geq& e^{-n( \min_{\pi_{UV}}D(\pi_{UV}\| Q_{UV} )+\eta_n)}.
\label{eq:bound_beta_s}
\end{IEEEeqnarray}

As a consequence: 
\begin{IEEEeqnarray}{rCl}
\beta_n
& = & \Pr\left[ \hat{\mathcal{H}}=0 | \mathcal{H}=1\right]\\ & \geq & \Pr\left[ \hat{\mathcal{H}}=0 ,X^n= \bar{x}^n| \mathcal{H}=1\right]\\
& \geq & \sum_{s \in \mathcal{\bar{S}}_{\bar{x}^n}  }\Pr[S=s] \nonumber \\
&&\hspace{0.7cm}\cdot \  \Pr\left[ \hat{\mathcal{H}}=0 ,X^n= \bar{x}^n| \mathcal{H}=1, S=s\right] \\ 
& \geq & \Pr[S \in \mathcal{\bar{S}}_{\bar{x}^n}] \left( e^{-n\big(\min_{\pi_{UV}} D(\pi_{UV} \| Q_{UV}) +\eta_n\big)} \right),\IEEEeqnarraynumspace\label{eq:lower_beta} 
\end{IEEEeqnarray}
where the minimum is over all pmfs $\pi_{UV}$ with marginals $P_U$ and $P_V$.

We next bound the probability of the set ${s}\in \mathcal{\bar{S}}_{\bar{x}^n}$. To this end, notice
\begin{IEEEeqnarray}
    {rCl}
    \lefteqn{\Pr\left [ \hat{\mathcal{H}}=0, X^n=\bar{x}^n\Big | \mathcal{H}=0\right]}\nonumber \\
    &=& \sum_{s\in \mathcal{\bar{S}}_{\bar{x}^n}} \Pr\left[ \hat{\mathcal{H}}=0, X^n=x^n| \mathcal{H}=0, S=s\right]\Pr[S=s] \nonumber\\
    && +  \sum_{s\notin \mathcal{\bar{S}}_{\bar{x}^n}} \Pr\left[ \hat{\mathcal{H}}=0, X^n=x^n| \mathcal{H}=0, S=s\right] \Pr[S=s]\nonumber \\\\
    &\leq& \Pr\left[ S\in  \mathcal{\bar{S}}_{\bar{x}^n}\right] \nonumber \\
    && + \sum_{s\notin \Bar{ \mathcal{S}}_{\Bar{x}^n}} \Pr\left[ \hat{\mathcal{H}}=0, X^n=x^n| \mathcal{H}=0, S=s\right]  \Pr(S=s)\nonumber \\\\
    &\stackrel{(a)}{\leq}& \Pr\left[ S\in  \Bar{ \mathcal{S}}_{\Bar{x}^n}\right] + \frac{1- \epsilon-\eta}{c|\tilde{\mathcal{X}}_0^n|} \bigg(1- \Pr\left[ s\in  \Bar{ \mathcal{S}}_{\Bar{x}^n}\right]\bigg)\\
    &=& \Pr\left[ S\in  \Bar{ \mathcal{S}}_{\Bar{x}^n}\right] \Bigg( 1 - \frac{1- \epsilon-\eta}{c|\tilde{\mathcal{X}}_0^n|}\Bigg) +  \frac{1- \epsilon-\eta}{c|\tilde{\mathcal{X}}_0^n|},
    \label{eq:bound_1}
\end{IEEEeqnarray}
where $(a)$ holds by  the definition of $\bar{S}_{\bar{x}^n}$.
Combining \eqref{eq:bound} and \eqref{eq:bound_1}, we then obtain: 
\begin{IEEEeqnarray}{rCl}
\Pr\left[ S\in  \Bar{ \mathcal{S}}_{\Bar{x}^n}\right]
&\geq&\frac{1- \epsilon-\eta}{|\tilde{\mathcal{X}}_0^n|} \bigg(1- \frac{1}{c}\bigg) \Bigg( 1 - \frac{1- \epsilon-\eta}{c|\tilde{\mathcal{X}}_0^n|}\Bigg)^{-1}  \IEEEeqnarraynumspace\\
&\geq& \frac{1- \epsilon-\eta}{|\tilde{\mathcal{X}}_0^n|} \bigg(1- \frac{1}{c}\bigg).
\label{eq:bound_p_s}
\end{IEEEeqnarray}

The proof is concluded by combining \eqref{eq:lower_beta} with \eqref{eq:bound_p_s} and \eqref{eq:sizeX}, and by letting $n\to \infty$:
\begin{IEEEeqnarray}{rCl}
\lefteqn{
\varlimsup_{n\to \infty} -\frac{1}{n} \log \beta_n} \quad \\ & \leq & \hspace{-0.1cm}\varlimsup_{n\to \infty} -\frac{1}{n} \log \Pr[S \in \Bar{ \mathcal{S}}_{\Bar{x}^n}] + \min_{\pi_{UV}} D(\pi_{UV} \| Q_{UV}) \IEEEeqnarraynumspace\\
& = & \min_{\pi_{UV}} D(\pi_{UV} \| Q_{UV}).
\label{eq:achi_part}
\end{IEEEeqnarray}
where the two minima are over all types $\pi_{UV}$ with marginals $P_U$ and $P_V$.

This proves that  $\theta_{\textnormal{covert},\mathcal H=0}^{*}(\epsilon)$ and  $\theta_{\textnormal{covert,both}}^{*}(\epsilon)$ are upper bounded by $E_1$.

\subsection{Upper Bound $E_2$}

Recall the definition of $\widetilde{\mathcal{US}}_0$ as the set of $(u^n,s)$ pairs mapping to low-weight inputs in \eqref{eq:US}. By Lemma~\ref{lem:lemma_low_weight}:
\begin{IEEEeqnarray}{rCl}\label{eq:prob1_key}
\lim_{n\to\infty}
\Pr\bigl[(U^n,S) \notin\widetilde{\mathcal{US}}_0\mid \mathcal{H}=0\bigr]
&=&0.
\end{IEEEeqnarray}

Let $\mu_n>0$ satisfy
\begin{subequations}\label{eq:mu_eq_key}
\begin{IEEEeqnarray}{rCl}
\lim_{n\to\infty}\mu_n&=&0,\\
\lim_{n\to\infty}n\mu_n^2&=&\infty.
\end{IEEEeqnarray}
\end{subequations}

Next, fix $\eta\in(0,1-\epsilon)$, and define the set
\begin{IEEEeqnarray}{rCl}
\mathcal D_n
&\triangleq&
\bigl\{(u^n,v^n,s):
 (u^n,s)\in\widetilde{\mathcal{US}}_0,
\ v^n\in\mathcal T_{\mu_n}^{(n)}(P_V),\nonumber\\
&&\hspace{-0.1cm}
\Pr\bigl[Y^n\in\mathcal A_n(v^n,s)
\mid U^n=u^n,V^n=v^n,S=s\bigr]\ge\eta\bigr\}. \nonumber
\\
\label{eq:dn_final}
\end{IEEEeqnarray}
The last conditional probability in \eqref{eq:dn_final} is the same under both hypotheses, since it equals
\begin{equation}
\Gamma_{Y|X}^{\otimes n}
\bigl(\mathcal A_n(v^n,s)\mid f^{(n)}(u^n,s)\bigr),
\end{equation}
and hence depends only on the encoding function and the channel transition law.

Notice that 
\begin{IEEEeqnarray}{rCl}
\Delta_n&\triangleq&\Pr\left[(U^n,V^n,S)\in\mathcal D_n\mid\mathcal H=0\right]\nonumber\\
&\geq&1-\Pr\left[(U^n,S)\notin\widetilde{\mathcal{US}}_0\mid\mathcal H=0\right]\nonumber\\
&&-\Pr\left[V^n\notin\mathcal T_{\mu_n}^{(n)}(P_V)\mid\mathcal H=0\right]\nonumber\\
&&-\Pr\biggl[\Pr\Bigl[Y^n\in\mathcal A_n(V^n,S)\mid U^n,V^n,S\Bigr]<\eta\;\Bigm|\;\mathcal H=0\biggr]\nonumber\\
&\geq&1-\Pr\left[(U^n,S)\notin\widetilde{\mathcal{US}}_0\mid\mathcal H=0\right]\nonumber\\
&&-\frac{|\mathcal V|}{4n\mu_n^2}-\frac{\epsilon}{1-\eta},
\label{eq:d_n_0_bound_key}
\end{IEEEeqnarray}
where the last inequality holds by Chebyshev's inequality and the union bound:
\begin{IEEEeqnarray}{rCl}
\lefteqn{\Pr\left[V^n\notin\mathcal T_{\mu_n}^{(n)}(P_V)\mid\mathcal H=0\right]}\nonumber\\
&\leq&\sum_{v\in\mathcal V}\Pr\left[\left|\pi_{V^n}(v)-P_V(v)\right|>\mu_n\mid\mathcal H=0\right]\nonumber\\
&\leq&\frac{|\mathcal V|}{4n\mu_n^2},
\end{IEEEeqnarray}
as well as Markov's inequality:
\begin{IEEEeqnarray}{rCl}
\lefteqn{\Pr\left[\Pr\left[Y^n\in\mathcal A_n(V^n,S)\mid U^n,V^n,S\right]<\eta\mid\mathcal H=0\right]}\nonumber\\
&=&\Pr\left[\Pr\left[Y^n\notin\mathcal A_n(V^n,S)\mid U^n,V^n,S\right]\geq 1-\eta\mid\mathcal H=0\right]\nonumber\\
&\leq&\frac{\mathbb E\left[\Pr\left[Y^n\notin\mathcal A_n(V^n,S)\mid U^n,V^n,S\right] \mid\mathcal H=0\right]}{1-\eta}\nonumber\\
&\leq&\frac{\epsilon}{1-\eta}. 
\end{IEEEeqnarray}
%where the last inequality follows from the Type-I error constraint.
Consequently, by Lemma~\ref{lem:lemma_low_weight} and the assumptions on $\mu_n$, we obtain
\begin{equation}
\varliminf_{n\to\infty}\Delta_n\geq 1-\frac{\epsilon}{1-\eta}>0.
\end{equation}

For each $(u^n,s)$, define the set of indices at which the encoder produces input zero:
\begin{equation}
\mathcal I_{u^n,s}
\triangleq
\bigl\{i\in\{1,\ldots,n\}: f_i^{(n)}(u^n,s)=0\bigr\},
\end{equation}
where $f_i^{(n)}(u^n,s)$ is the $i$-th component of $f^{(n)}(u^n,s)$. %We also write $I\triangleq\mathcal I_{U^n,S}$.
To prove the desired converse result, we define the new pmf
\begin{IEEEeqnarray}{rCl}
\lefteqn{\widetilde P_{U^nV^nSY^n \mathcal I_{u^n,s}}(u^n,v^n,s,y^n,\mathcal I)}\nonumber\\
&=&\frac{P_{UV}^{\otimes n}(u^n,v^n)P_S(s)}{\Delta_n}
\,\mathbbm 1\bigl\{(u^n,v^n,s)\in\mathcal D_n\bigr\}\nonumber\\
&&\quad\cdot\Gamma_{Y|X}^{\otimes n}
\bigl(y^n\mid f^{(n)}(u^n,s)\bigr)
\,\mathbbm 1\bigl\{\mathcal I=\mathcal I_{u^n,s}\bigr\}.
\label{eq:tilted_measure_key}
\end{IEEEeqnarray}
This is the original law under $\mathcal H_0$, conditioned on
$(U^n,V^n,S)\in\mathcal D_n$.

By the definition of $\mathcal D_n$, for every $(u^n,v^n,s)\in\mathcal D_n$, we have
\begin{IEEEeqnarray}{rCl}
\Pr\left[Y^n\in\mathcal A_n(v^n,s)\mid U^n=u^n,V^n=v^n,S=s\right]&\geq&\eta.\nonumber \\
\end{IEEEeqnarray}
Applying the blowing-up lemma \cite[Remark, p.~446]{MartonBU} to $\mathcal A_n(v^n,s)$, there exists a  sequence of positive integers $\{\ell_n\}$ such that $\ell_n/\sqrt{n}\to\infty$ and $\ell_n/n\to0$, and 
\begin{IEEEeqnarray}{rCl}
\Pr\left[Y^n\in\widehat{\mathcal A}_n^{\ell_n}(v^n,s)\mid U^n=u^n,V^n=v^n,S=s\right]\geq 1-\lambda_n, \nonumber \\
\label{eq:blow_uplemma}
\end{IEEEeqnarray}
where $\lambda_n\triangleq\sqrt{n\ln(1/\eta)}/\ell_n\to0$ as $n\to\infty$.

Therefore, the new measure defined in \eqref{eq:tilted_measure_key} satisfies:
\begin{IEEEeqnarray}{rCl}
\lefteqn{\widetilde P\left[Y^n\in\widehat{\mathcal A}_n^{\ell_n}(V^n,S)\right]}\nonumber\\
&=&\sum_{(u^n,v^n,s)\in\mathcal D_n}\widetilde P_{U^nV^nS}(u^n,v^n,s)\nonumber\\
&&\hspace{1cm}\cdot\Pr\left[Y^n\in\widehat{\mathcal A}_n^{\ell_n}(v^n,s)\mid U^n=u^n,V^n=v^n,S=s\right]\nonumber\\
&\stackrel{(a)}{\geq}&1-\lambda_n,
\label{eq:P0}
\end{IEEEeqnarray}
where Inequality $(a)$ follows by \eqref{eq:blow_uplemma}.

We next examine how the blow-up of the acceptance region influences the Type-II error probability:
\begin{IEEEeqnarray}{rCl}
\lefteqn{\Pr\left[\left(Y^n\in\widehat{\mathcal A}_n^{\ell_n}(V^n,S)\right)\big| \mathcal{H}=1\right]}\nonumber\\
&=&\sum_{(u^n,v^n,s)}Q_{UV}^{\otimes n}(u^n,v^n)P_S(s)\nonumber\\
&&\hspace{.2cm}\cdot\Pr\left[Y^n\in\widehat{\mathcal A}_n^{\ell_n}(v^n,s)\mid U^n=u^n,V^n=v^n,S=s, \mathcal{H}=1\right]\nonumber\\
&\stackrel{(b)}{\leq}&\sum_{(u^n,v^n,s)}Q_{UV}^{\otimes n}(u^n,v^n)P_S(s)\nonumber\\
&&\hspace{.2cm}\cdot\Pr\left[Y^n\in\mathcal A_n(v^n,s)\mid U^n=u^n,V^n=v^n,S=s,\mathcal{H}=1\right]\nonumber\\
&&\hspace{.2cm}\cdot  e^{nH_{\textnormal b}(\ell_n/n)}|\mathcal Y|^{\ell_n}\gamma_{\min}^{-\ell_n}\label{step62}\\
&=&\beta_n e^{nH_{\textnormal b}(\ell_n/n)}|\mathcal Y|^{\ell_n}\gamma_{\min}^{-\ell_n},
\label{step64}
\end{IEEEeqnarray}
where $\gamma_{\min}$ is defined in \eqref{eq:E3_gamma_min}
 and Inequality $(b)$ follows from
\eqref{eq:blowup_probability_bound}.

\medskip
For ease of notation, let $Q_{V^nY^nS}$ denote the probability law under $\mathcal{H}=1$. Define
\begin{IEEEeqnarray}{rCl}
\widetilde P_{\mathcal H}(0)&\triangleq&\widetilde P\left(Y^n\in\widehat{\mathcal A}_n^{\ell_n}(V^n,S)\right),\\
Q_{\mathcal H}(0)&\triangleq&Q\left(Y^n\in\widehat{\mathcal A}_n^{\ell_n}(V^n,S)\right),
\end{IEEEeqnarray}
and let $\widetilde P_{\mathcal H}(1)$ and $Q_{\mathcal H}(1)$ denote their complements. Then, by data processing:
\begin{IEEEeqnarray}{rCl}
\lefteqn{D\left(\widetilde P_{V^nY^nS}\|Q_{V^nY^nS}\right) } \quad  \nonumber \\
&\geq&D\left(\widetilde P_{\mathcal H}\|Q_{\mathcal H}\right)\\
&=&-H_b (\widetilde P_{\mathcal H}(0))+\widetilde P_{\mathcal H}(0)\log\frac{1}{Q_{\mathcal H}(0)} \\
&\geq&-\log 2+\widetilde P_{\mathcal H}(0)\log\frac{1}{Q_{\mathcal H}(0)}.
\label{eq:D}
\end{IEEEeqnarray}
%where the first inequality follows from the data processing inequality and the last inequality follows from the bound $H_{\textnormal b}(p)\leq\log 2$.

We are now ready to bound the Type-II error probability.
Combining \eqref{eq:P0}, \eqref{step64}, and \eqref{eq:D}, we obtain
\begin{IEEEeqnarray}{rCl}
\lefteqn{-\frac{1}{n}\log\beta_n}\nonumber\\
&\leq&\frac{\log 2+D\left(\widetilde P_{V^nY^nS}\|Q_{V^nY^nS}\right)}{n(1-\lambda_n)}\nonumber\\
&&+H_{\textnormal b}(\ell_n/n)+\frac{\ell_n}{n}\log\frac{|\mathcal Y|}{\gamma_{\min}}\nonumber\\
&=&\frac{\log 2+D\left(\widetilde P_{V^nS}\|Q_V^{\otimes n}P_S\right)}{n(1-\lambda_n)}\nonumber\\
&&+\frac{1}{n(1-\lambda_n)}\mathbb E_{\widetilde P_{V^nS}}\left[D\left(\widetilde P_{Y^n|V^nS}\|Q_{Y^n|V^nS}\right)\right]\nonumber\\
&&+H_{\textnormal b}(\ell_n/n)+\frac{\ell_n}{n}\log\frac{|\mathcal Y|}{\gamma_{\min}}.
\label{eq:summands}
\end{IEEEeqnarray}

\medskip
We bound the two divergence terms in \eqref{eq:summands} individually. For the first divergence term we have:
\begin{IEEEeqnarray}{rCl}
\lefteqn{ \frac{1}{n} D\left(\tilde P_{V^nS}\|Q_V^{\otimes n}P_S\right)}\nonumber\\
&=&\frac{1}{n}\sum_{v^n,s}\tilde P_{V^nS}(v^n,s)\log\frac{\tilde P_{V^nS}(v^n,s)}{Q_V^{\otimes n}(v^n)P_S(s)}\nonumber\\
&\stackrel{(c)}{\leq}&\frac{1}{n} \sum_{v^n,s}\tilde P_{V^nS}(v^n,s)\log\frac{P_V^{\otimes n}(v^n)P_S(s)}{\Delta_n Q_V^{\otimes n}(v^n)P_S(s)}\nonumber\\
&=&\frac{1}{n} \sum_{v^n,s}\tilde P_{V^nS}(v^n,s)\log\frac{P_V^{\otimes n}(v^n)}{Q_V^{\otimes n}(v^n)}+\frac{1}{n} \log\frac{1}{\Delta_n}\nonumber\\
&\stackrel{(d)}{\leq}&D(P_V\|Q_V)+\mu_n\sum_{v}\left|\log\frac{P_V(v)}{Q_V(v)}\right|+ \frac{1}{n}\log\frac{1}{\Delta_n}, \nonumber \\
\label{eq:DV}
\end{IEEEeqnarray}
which tends to $D(P_V\| Q_V)$ because  $\mu_n\to0$ and $\left|\log\frac{P_V(v)}{Q_V(v)}\right|$ is bounded and because $\Delta_n$ is bounded away from 0 and smaller than~1. 
In the above, Inequality~ $(c)$ holds because
\begin{IEEEeqnarray}{rCl}
\tilde P_{V^nS}(v^n,s)
&=&\sum_{u^n}\tilde P_{U^nV^nS}(u^n,v^n,s)\\
&=&\frac{P_S(s)}{\Delta_n}\sum_{u^n}P_{UV}^{\otimes n}(u^n,v^n)\mathbbm{1}\{(u^n,v^n,s)\in\mathcal D_n\} \nonumber \\\\
&\leq&\frac{1}{\Delta_n}P_V^{\otimes n}(v^n)P_S(s),
\end{IEEEeqnarray}
and Inequality $(d)$ because $V^n\in\mathcal T_{\mu_n}^{(n)}(P_V)$.

\medskip
Consider next the second divergence term in \eqref{eq:summands}. Recall that $\mathcal I_{U^n,S}\triangleq\{i:f_i^{(n)}(U^n,S)=0\}$, and observe
\begin{IEEEeqnarray}{rCl}
\lefteqn{\tilde P_{Y^n|V^nS}(y^n|v^n,s)} \nonumber \\
&=&\sum_{\mathcal I}\tilde P_{\mathcal I_{U^n,S}|V^nS}(\mathcal I|v^n,s)\tilde P_{Y^n|V^nS\mathcal I_{U^n,S}}(y^n|v^n,s,\mathcal I)\nonumber\\
&=&\mathbb E_{\tilde P_{\mathcal I_{U^n,S}|V^n=v^n,S=s}}\left[\tilde P_{Y^n|V^nS\mathcal I_{U^n,S}}(y^n|v^n,s,\mathcal I_{U^n,S})\right]. \nonumber \\
\end{IEEEeqnarray}
%Similarly,
%\begin{IEEEeqnarray}{rCl}
%\lefteqn{Q_{Y^n|V^nS}(y^n|v^n,s)} \nonumber \\
%&=&\sum_{\mathcal I}\tilde P_{\mathcal I_{U^n,S}|V^nS}(\mathcal I|v^n,s)Q_{Y^n|V^nS}%(y^n|v^n,s)\nonumber\\
%&=&\mathbb E_{\tilde P_{\mathcal I_{U^n,S}|V^n=v^n,S=s}}\left[Q_{Y^n|V^nS}(y^n|v^n,s)\right].
%\end{IEEEeqnarray}
Using the convexity of the KL divergence in the first argument, we obtain
\begin{IEEEeqnarray}{rCl}
\lefteqn{\frac{1}{n}\mathbb E_{\tilde P_{V^nS}}\left[D\left(\tilde P_{Y^n|V^nS}\|Q_{Y^n|V^nS}\right)\right]}\nonumber\\
&\leq&\frac{1}{n}\mathbb E_{\tilde P_{V^nS\mathcal I_{U^n,S}}}\left[D\left(\tilde P_{Y^n|V^nS\mathcal I_{U^n,S}}\|Q_{Y^n|V^nS}\right)\right].\label{eq:conv}\IEEEeqnarraynumspace
\end{IEEEeqnarray}

For a given realization $\mathcal I_{U^n,S}=\mathcal I$, let $Y_{\mathcal I}^n\triangleq(Y_i:i\in\mathcal I)$ and $Y_{\mathcal I^c}^n\triangleq(Y_i:i\in\mathcal I^c)$ denote the subvectors of $Y^n$ corresponding to the zero and nonzero input positions, respectively, where $\mathcal I^c\triangleq\{1,\ldots,n\}\setminus\mathcal I$. We then decompose the divergence using the chain rule. Using obvious abbreviations:
\begin{IEEEeqnarray}{rCl}
\lefteqn{\frac{1}{n}D\left(\tilde P_{Y^n|V^n=v^n,S=s,\mathcal I_{U^n,S}=\mathcal I}\|Q_{Y^n|V^n=v^n,S=s}\right)}\nonumber\\
&=&\frac{1}{n}D\left(\tilde P_{Y_{\mathcal I}^n|v^n,s,\mathcal I}\|Q_{Y_{\mathcal I}^n|v^n,s}\right)\nonumber\\
&&+\frac{1}{n}\mathbb E_{\tilde P_{Y_{\mathcal I}^n|v^n,s,\mathcal I}}\left[D\left(\tilde P_{Y_{\mathcal I^c}^n|Y_{\mathcal I}^n,v^n,s,\mathcal I}\|Q_{Y_{\mathcal I^c}^n|Y_{\mathcal I}^n,v^n,s}\right)\right]\nonumber\\
&=&\frac{1}{n}D\left(\tilde P_{Y_{\mathcal I}^n|v^n,s,\mathcal I}\|Q_{Y_{\mathcal I}^n|v^n,s}\right)\nonumber\\
&&+\frac{1}{n}\sum_{y_{\mathcal I}^n}\tilde P_{Y_{\mathcal I}^n|v^n,s,\mathcal I}(y_{\mathcal I}^n)\sum_{y_{\mathcal I^c}^n}\tilde P_{Y_{\mathcal I^c}^n|y_{\mathcal I}^n,v^n,s,\mathcal I}(y_{\mathcal I^c}^n)\nonumber\\
&&\quad\cdot\log\frac{\tilde P_{Y_{\mathcal I^c}^n|y_{\mathcal I}^n,v^n,s,\mathcal I}(y_{\mathcal I^c}^n)}{Q_{Y_{\mathcal I^c}^n|y_{\mathcal I}^n,v^n,s}(y_{\mathcal I^c}^n)}\nonumber\\
&\leq&\frac{1}{n}D\left(\tilde P_{Y_{\mathcal I}^n|v^n,s,\mathcal I}\|Q_{Y_{\mathcal I}^n|v^n,s}\right)\nonumber\\
&&-\frac{1}{n}\sum_{y_{\mathcal I}^n}\tilde P_{Y_{\mathcal I}^n|v^n,s,\mathcal I}(y_{\mathcal I}^n)\sum_{y_{\mathcal I^c}^n}\tilde P_{Y_{\mathcal I^c}^n|y_{\mathcal I}^n,v^n,s,\mathcal I}(y_{\mathcal I^c}^n)\nonumber\\
&&\quad\cdot\log Q_{Y_{\mathcal I^c}^n|y_{\mathcal I}^n,v^n,s}(y_{\mathcal I^c}^n).
\label{kl_decomp_0}
\end{IEEEeqnarray}
The last inequality holds because $\log\tilde P_{Y_{\mathcal I^c}^n|y_{\mathcal I}^n,v^n,s,\mathcal I}(y_{\mathcal I^c}^n)\leq0$.

Continue to notice that since the channel is fully-connected:
\begin{IEEEeqnarray}{rCl}
\lefteqn{Q_{Y_{\mathcal I^c}^n|Y_{\mathcal I}^n,V^n,S}(y_{\mathcal I^c}^n|y_{\mathcal I}^n,v^n,s)}\nonumber\\
&=&\sum_{x^n\in\mathcal X^n}Q_{X^n|Y_{\mathcal I}^n,V^n,S}(x^n|y_{\mathcal I}^n,v^n,s)\prod_{i\in\mathcal I^c}\Gamma_{Y|X}(y_i|x_i)\nonumber\\
&\geq&\sum_{x^n\in\mathcal X^n}Q_{X^n|Y_{\mathcal I}^n,V^n,S}(x^n|y_{\mathcal I}^n,v^n,s)\gamma_{\min}^{n-|\mathcal I|}\nonumber\\
&=&\gamma_{\min}^{n-|\mathcal I|}.
\end{IEEEeqnarray}
%Here, the inequality follows from $\Gamma_{Y|X}(y|x)\geq\gamma_{\min}>0$ for all $(x,y)$.
%
Substituting this bound into \eqref{kl_decomp_0}, we obtain
\begin{IEEEeqnarray}{rCl}
\lefteqn{\frac{1}{n}D\left(\tilde P_{Y^n|v^n,s,\mathcal I}\|Q_{Y^n|v^n,s}\right)}\nonumber\\
&\leq&\frac{1}{n}D\left(\tilde P_{Y_{\mathcal I}^n|v^n,s,\mathcal I}\|Q_{Y_{\mathcal I}^n|v^n,s}\right)-\frac{n-|\mathcal I|}{n}\log\gamma_{\min}.\IEEEeqnarraynumspace
\label{eq:div_comp}
\end{IEEEeqnarray}

Under $\tilde P$, for any $(v^n,s,\mathcal I)$ in the support of $\tilde P_{V^nS\mathcal I_{U^n,S}}$, we have
\begin{IEEEeqnarray}{rCl}
\tilde P_{Y_{\mathcal I}^n|V^n=v^n,S=s,\mathcal I_{U^n,S}=\mathcal I}
&=&\Gamma_{Y|X}(\cdot|0)^{\otimes|\mathcal I|},
\end{IEEEeqnarray}
because $f_i^{(n)}(u^n,s)=0$ for every $i\in\mathcal I$, and the channel is memoryless.

Therefore, from \eqref{eq:div_comp}, we obtain
\begin{IEEEeqnarray}{rCl}
\lefteqn{\frac{1}{n}D\left(\tilde P_{Y^n|v^n,s,\mathcal I}\|Q_{Y^n|v^n,s}\right)}\nonumber\\
&\leq&\frac{1}{n}D\left(\Gamma_{Y|X}(\cdot|0)^{\otimes|\mathcal I|}\|Q_{Y_{\mathcal I}^n|v^n,s}\right)-\frac{n-|\mathcal I|}{n}\log\gamma_{\min}. \nonumber \\
\label{eq:ne}
\end{IEEEeqnarray}

\medskip
We continue to bound the divergence term in \eqref{eq:ne}. Notice that 
\begin{equation}
Q_{Y_{\mathcal I}^n|v^n,s}=\sum_{x^n}Q_{X^n|v^n,s}(x^n)\prod_{i\in\mathcal I}\Gamma_{Y|X}(\cdot|x_i),
\end{equation}
and by the convexity of the KL divergence in its second argument: 
\begin{IEEEeqnarray}{rCl}
\lefteqn{\frac{1}{n}D\left(\Gamma_{Y|X}(\cdot|0)^{\otimes|\mathcal I|}\|Q_{Y_{\mathcal I}^n|v^n,s}\right)}\nonumber\\
&\leq&\frac{1}{n}\sum_{x^n}Q_{X^n|v^n,s}(x^n)D\left(\Gamma_{Y|X}(\cdot|0)^{\otimes|\mathcal I|}\middle\|\prod_{i\in\mathcal I}\Gamma_{Y|X}(\cdot|x_i)\right)\nonumber\\
&=&\frac{1}{n}\sum_{x^n}Q_{X^n|v^n,s}(x^n)\sum_{i\in\mathcal I}D\left(\Gamma_{Y|X}(\cdot|0)\|\Gamma_{Y|X}(\cdot|x_i)\right)\nonumber\\
&\leq&\frac{|\mathcal I|}{n}\max_{x'\in\mathcal X}D\left(\Gamma_{Y|X}(\cdot|0)\|\Gamma_{Y|X}(\cdot|x')\right)\nonumber\\
&\leq&\max_{x'\in\mathcal X}D\left(\Gamma_{Y|X}(\cdot|0)\|\Gamma_{Y|X}(\cdot|x')\right).
\label{eq:div_type_fin}
\end{IEEEeqnarray}
%Here, the equality follows from the additivity of the KL divergence for product distributions. The first inequality after the equality follows by upper-bounding each divergence by its maximum over $\mathcal X$, while the last inequality follows from $|\mathcal I|\leq n$.

Substituting \eqref{eq:div_type_fin} into \eqref{eq:ne} yields
\begin{IEEEeqnarray}{rCl}
\lefteqn{\frac{1}{n}D\left(\tilde P_{Y^n|v^n,s,\mathcal I}\|Q_{Y^n|v^n,s}\right)}\nonumber\\
&\leq&\max_{x'\in\mathcal X}D\left(\Gamma_{Y|X}(\cdot|0)\|\Gamma_{Y|X}(\cdot|x')\right)-\frac{n-|\mathcal I|}{n}\log\gamma_{\min}. \nonumber \\
\label{eq:div_final}
\end{IEEEeqnarray}

To bound the last term in \eqref{eq:div_final}, notice that, by the definition of $\mathcal D_n$, under $\tilde P$ we have
\begin{equation}
n-|\mathcal I_{U^n,S}|=w_H(f^{(n)}(U^n,S))\leq \bar{a}_n^0 \cdot n,\label{eq:bounda}
\end{equation}
where $\bar{a}_n^0$ is defined in \eqref{eq:eps_n} and vanishes as $n\to \infty$, see \eqref{eq:van_a0}. 
%Since $k'(n)=o(n)$ and $\gamma_{\min}>0$, % the second term in \eqref{eq:div_final} vanishes. Combining \eqref{eq:div_final} with the convexity bound gives
Combining Inequality \eqref{eq:bounda} with Inequalities \eqref{eq:conv} and  \eqref{eq:div_final} yields
\begin{IEEEeqnarray}{rCl}
\lefteqn{\frac{1}{n}\mathbb E_{\tilde P_{V^nS}}\left[D\left(\tilde P_{Y^n|V^nS}\|Q_{Y^n|V^nS}\right)\right]}\quad \nonumber\\
&\leq&\max_{x'\in\mathcal X}D\left(\Gamma_{Y|X}(\cdot|0)\|\Gamma_{Y|X}(\cdot|x')\right)-\bar{a}_n^0\log\gamma_{\min}. \IEEEeqnarraynumspace
\label{eq:second_term_bound}
\end{IEEEeqnarray}

Finally, combining \eqref{eq:second_term_bound} with \eqref{eq:summands} and \eqref{eq:DV}, and using the facts that $\ell_n/n\to0$, $\mu_n\to0$, $\gamma_{\min}>0$, $\bar a_n^0\to 0$, and that $\Delta_n$ is bounded away from 0 and smaller than 1, we  obtain the desired inequality:
\begin{IEEEeqnarray}{rCl}
\varlimsup_{n\to\infty}-\frac{1}{n}\log\beta_n
&\leq&D(P_V\|Q_V)\nonumber\\
&&+\max_{x'\in\mathcal X}D\left(\Gamma_{Y|X}(\cdot|0)\|\Gamma_{Y|X}(\cdot|x')\right). \IEEEeqnarraynumspace
\end{IEEEeqnarray}

\subsection{Upper Bound $E_3$}
Let $F_{UV}$ be an arbitrary pmf satisfying $F_U=Q_U$ and $F_V=P_V$. Throughout this proof, $\Pr_{F}[\cdot]$ denotes probability assuming $(U^n,V^n) \sim F_{UV}^{\otimes n}$. 

\newcommand{\f}{\zeta}
Recall the definition of $\widetilde{\mathcal US}_{\textnormal{both}}$ in \eqref{eq:USboth}, fix a vanishing sequence $\mu_n\to 0$ as $n\to \infty$, and define the region
\begin{IEEEeqnarray}{rCl}
\mathcal F_n&\triangleq&\{(u^n,v^n,s,y^n):(u^n,v^n)\in\mathcal T_{\mu_n}^{(n)}(F_{UV}),\nonumber\\
&&\hspace{0.7cm}(u^n,s)\in\widetilde{\mathcal US}_{\textnormal{both}},\;g^{(n)}(v^n,y^n,s)=0\},
\label{eq:E3_Fn}
\end{IEEEeqnarray}
and its probability under $F_{UV}^{\otimes n}$:
\begin{IEEEeqnarray}{rCl}
\f_n&\triangleq&{\Pr}_{F}[(U^n,V^n,S,Y^n)\in\mathcal F_n].
\label{eq:E3_Delta}
\end{IEEEeqnarray}

Start by bounding: 
\begin{IEEEeqnarray}{rCl}
\beta_n&=&\Pr[g^{(n)}(V^n,Y^n,S)=0|\mathcal H=1] \nonumber \\
&\ge& \Pr[(U^n,V^n,S,Y^n)\in\mathcal F_n|\mathcal H=1] \nonumber \\
&=&
\sum_{\substack{(u^n,v^n)\in\mathcal T_{\mu_n}^{(n)}(F_{UV})\\
(u^n,s)\in\widetilde{\mathcal US}_{\textnormal{both}}}}
Q_{UV}^{\otimes n}(u^n,v^n)P_S(s)\nonumber\\
&&\quad\cdot\Pr[Y^n\in\mathcal A_n(v^n,s)|X^n=f^{(n)}(u^n,s)]\nonumber\\
&\stackrel{(a)}{\geq}& e^{-n(D(F_{UV}\|Q_{UV})+\delta_n)}
\hspace{-0.7cm}\sum_{\substack{(u^n,v^n)\in\mathcal T_{\mu_n}^{(n)}(F_{UV})\\
(u^n,s)\in\widetilde{\mathcal US}_{\textnormal{both}}}}
F_{UV}^{\otimes n}(u^n,v^n)P_S(s)\nonumber\\
&&\quad\cdot\Pr[Y^n\in\mathcal A_n(v^n,s)|X^n=f^{(n)}(u^n,s)]\nonumber\\
&=&e^{-n(D(F_{UV}\|Q_{UV})+\delta_n)} \cdot \f_n,
\label{eq:E3_beta_lower_bound}
\end{IEEEeqnarray}
where  $\delta_n$ is a sequence that vanishes with $\mu_n$. Inequality $(a)$ holds because 
for any $(u^n,v^n)\in\mathcal T_{\mu_n}^{(n)}(F_{UV})$, we have
\begin{IEEEeqnarray}{rCl}
\lefteqn{\frac1n\log\frac{F_{UV}^{\otimes n}(u^n,v^n)}{Q_{UV}^{\otimes n}(u^n,v^n)}}\nonumber\\
%&=&\frac1n\sum_{i=1}^n\log\frac{F_{UV}(u_i,v_i)}{Q_{UV}(u_i,v_i)}\nonumber\\
&=&\sum_{(u,v)}\pi_{u^nv^n}(u,v)\log\frac{F_{UV}(u,v)}{Q_{UV}(u,v)}\\
&=&D(F_{UV}\|Q_{UV}) \nonumber \\
&+&\sum_{(u,v)}\left(\pi_{u^nv^n}(u,v)-F_{UV}(u,v)\right)\log\frac{F_{UV}(u,v)}{Q_{UV}(u,v)}\nonumber\\
&\leq&D(F_{UV}\|Q_{UV})+\sum_{(u,v)}\left|\pi_{u^nv^n}(u,v)-F_{UV}(u,v)\right| \nonumber \\
&&\hspace{4cm}\left|\log\frac{F_{UV}(u,v)}{Q_{UV}(u,v)}\right|\nonumber\\
&\leq&D(F_{UV}\|Q_{UV})+\mu_n\sum_{(u,v)}\left|\log\frac{F_{UV}(u,v)}{Q_{UV}(u,v)}\right|.
\label{eq:E3_likelihood_ratio}
\end{IEEEeqnarray}

Since $F_{UV}$ was chosen arbitrary with marginals $Q_U$ and $P_V$, the bound $E_3$ is established by \eqref{eq:E3_beta_lower_bound} if we can show that 
\begin{equation} \label{eq:f}
\lim_{n\to \infty} \frac{1}{n} \log \f_n =0.
\end{equation} 

Let 
\begin{IEEEeqnarray}{rCl}
k'&\triangleq&n \cdot\max\left\{\bar a_n^{(0)},\bar a_n^{(1)}\right\}.
\label{eq:E3_kprime}
\end{IEEEeqnarray}
By Lemma~\ref{lem:lemma_low_weight},  $\bar a_n^{(0)}\rightarrow 0$ and $\bar a_n^{(1)}\rightarrow0$, and thus
\begin{IEEEeqnarray}{rCl}
\lim_{n\to\infty}\frac{k'}n&=&0,
\label{eq:E3_kprime_limit}
\end{IEEEeqnarray}
and moreover for $H\in\{0,1\}$:
\begin{IEEEeqnarray}{rCl}
\lim_{n\to \infty}\Pr[(U^n,S)\notin\widetilde{\mathcal US}_{\textnormal{both}}|\mathcal H=H]=0.\IEEEeqnarraynumspace
\label{eq:E3_low_weight}
\end{IEEEeqnarray}

\begin{comment}

By the definition of $\mathcal F_n$, we have
\begin{IEEEeqnarray}{rCl}
\Delta_n&=&\Pr[(U^n,S)\in\widetilde{\mathcal{US}}^{n}]\nonumber\\
&&\cdot\Pr[(U^n,V^n)\in\mathcal T_{\mu_n}^{(n)}(F_{UV})|(U^n,S)\in\widetilde{\mathcal US}^{n}]\nonumber\\
&&\cdot\Pr[g^{(n)}(V^n,Y^n,S)=0|(U^n,S)\in\widetilde{\mathcal{US}}^{n},\nonumber \\
&& \hspace{3.3cm}(U^n,V^n)\in\mathcal T_{\mu_n}^{(n)}(F_{UV})]. \nonumber \\
\label{eq:E3_Delta_decomposition}
\end{IEEEeqnarray}
Notice that, since $F_U=Q_U$, by \eqref{eq:E3_low_weight},
\begin{IEEEeqnarray}{rCl}
\Pr[(U^n,S)\in\widetilde{\mathcal{US}}^{n}]&=&\Pr[(U^n,S)\in\widetilde{\mathcal{US}}^{n}|\mathcal H=1]  \nonumber \\
&\geq&1-o(1).
\label{eq:p_low_weigt_f}
\end{IEEEeqnarray}
Moreover, by the law of large numbers,
\begin{IEEEeqnarray}{rCl}
\Pr[(U^n,V^n)\in\mathcal T_{\mu_n}^{(n)}(F_{UV})]&\geq&1-o(1),
\label{eq:E3_typicality}
\end{IEEEeqnarray}
where $\mu_n\to0$.
\end{comment}

%Next, let $Y_0^n$ be an auxiliary random sequence generated according to $\Gamma_{Y|X=0}^{\otimes n}$, independently of $(U^n,V^n,S)$. We couple $Y^n$ and $Y_0^n$ as follows. For each $i\in\{1,\ldots,n\}$, if $X_i=0$, set $Y_i=Y_{0,i}$; otherwise, generate $Y_i$ according to $\Gamma_{Y|X}(\cdot|X_i)$, independently of $Y_{0,i}$. 

%By construction, 
%\begin{IEEEeqnarray}{rCl}
%d_{\mathrm H}(Y^n,Y_0^n)&\leq&w_{\mathrm H}(X^n) \\
%&\leq&k'(n).
%\label{eq:E3_coupling_low_weight}
%\end{IEEEeqnarray}

%\noindent\underline{\textit{Probability of the enlarged acceptance region:}}\\
Let $Y_0^n$ be a sequence drawn according to $\Gamma_{Y|X=0}^{\otimes n}$, independently of $(U^n,V^n,S)$.  Notice that when $(u^n,s)\in \widetilde{\mathcal{US}}_{\textnormal{both}}$, the generation of $Y_0^n$ differs from the generation of $Y^n$  only in $k'$ positions. We thefore  obtain the  following two inequalities: 
\begin{IEEEeqnarray}{rCl}
\lefteqn{{\Pr}_F\left[Y^n\in\mathcal A_n(V^n,S), \ (U^n,S)\in\widetilde{\mathcal{US}}_{\textnormal{both}} \right]} \quad \nonumber \\
& \leq & {\Pr}_F \left[ Y_0^n\in\mathcal A_n^{k'}(V^n,S) \right],
\label{eq:E3_event_inclusion}
\end{IEEEeqnarray}
and
\begin{IEEEeqnarray}{rCl}
\lefteqn{{\Pr}_F\left[Y_0^n\in\mathcal A_n^{k'}(V^n,S), \ (U^n,S)\in\widetilde{\mathcal{US}}_{\textnormal{both}} \right]} \quad \nonumber \\
& \leq & {\Pr}_F \left[ Y^n\in\mathcal A_n^{2k'}(V^n,S) \right],
\label{eq:E3_event_inclusion2}
\end{IEEEeqnarray}

%\begin{IEEEeqnarray}{rCl}
%\Pr[Y_0^n\in\mathcal A_n^{k'(n)}(V^n,S)]&=&\Pr[Y_0^n\in\mathcal A_n^{k'(n)}(V^n,S)|\mathcal H=0]. \nonumber \\
%\label{eq:E3_equal_marginals}
%\end{IEEEeqnarray}
%Under $\mathcal H=0$,\eqref{eq:E3_coupling_low_weight} implies

%Thus,
Using these inequalities and recalling that $F_V=P_V$, we   write: 
\begin{IEEEeqnarray}{rCl}
\lefteqn{ {\Pr}_F \left[ Y^n\in\mathcal A_n^{2k'}(V^n,S) \right]} \qquad \nonumber \\[1.2ex]
&\geq & {\Pr}_F[Y_0^n\in\mathcal A_n^{k'(n)}(V^n,S), \; (U^n,S)\in\widetilde{\mathcal{US}}_{\textnormal{both}}]\nonumber \\[1.2ex] 
&\geq  & {\Pr}_F[Y_0^n\in\mathcal A_n^{k'(n)}(V^n,S) ] -  {\Pr}_F[(U^n,S)\in\widetilde{\mathcal{US}}_{\textnormal{both}}]\nonumber \\
& \stackrel{(b)}{=} &\Pr[Y_0^n\in\mathcal A_n^{k'(n)}(V^n,S)|\mathcal H=0] -o(1) \nonumber\\[1.2ex]
&{\geq} &\Pr[Y^n\in\mathcal A_n(V^n,S),(U^n,S)\in\widetilde{\mathcal{US}}_{\textnormal{both}}|\mathcal H=0]\nonumber\\ && -o(1) \nonumber\\
&{\geq}&\Pr[Y^n\in\mathcal A_n(V^n,S)|\mathcal H=0]\nonumber\\
&&\quad-\Pr[(U^n,S)\notin\widetilde{\mathcal{US}}_{\textnormal{both}}|\mathcal H=0]-o(1) \nonumber\\
&\stackrel{(c)}{\geq}&1-\epsilon-o(1),
\label{eq:E3_enlarged_auxiliary}
\end{IEEEeqnarray}
where Equality $(b)$ holds by \eqref{eq:E3_low_weight} and  because $F_U=Q_U$; and Inequality $(c)$ holds  because the  Type-I error probability is upper bounded by $\epsilon$ and again by  \eqref{eq:E3_low_weight}. 

We are ready to lower bound $\f_n$ and prove \eqref{eq:f}. In fact, 
\begin{IEEEeqnarray}{rCl}
\f_n %&= & {\Pr}_{F}[(U^n,V^n,S,Y^n)\in\mathcal F_n]\\
& = & {\Pr}_F\big[Y^n\in\mathcal A_n(V^n,S), \;(U^n,S)\in\widetilde{\mathcal{US}}_{\textnormal{both}},\nonumber \\
&& \hspace{3cm}(U^n,V^n)\in\mathcal T_{\mu_n}^{(n)}(F_{UV})\big]\IEEEeqnarraynumspace\\
&\stackrel{(d)}{\geq }  &\frac{\gamma_{\min}^{2k'}}{B_n(2k')}  {\Pr}_F\big[Y^n\in\mathcal A^{2k'}_n(V^n,S),\;(U^n,S)\in\widetilde{\mathcal{US}}_{\textnormal{both}},\nonumber \\
&& \hspace{4.3cm}(U^n,V^n)\in\mathcal T_{\mu_n}^{(n)}(F_{UV}) 
\big]\nonumber\\
& \geq& \frac{\gamma_{\min}^{2k'}}{B_n(2k')} \Big({\Pr}_F[Y^n\in\mathcal A_n^{2k'}(V^n,S)]  \nonumber \\
&& \hspace{1.8cm} - {\Pr}_F[(U^n,V^n)\notin\mathcal T_{\mu_n}^{(n)}(F_{UV})]\nonumber\\
&& \hspace{2.8cm}-{\Pr}_F[(U^n,S)\notin\widetilde{\mathcal{US}}_{\textnormal{both}} ]\Big)\\
& \stackrel{(e)}{= } & \frac{\gamma_{\min}^{2k'}}{B_n(2k')} \left[{\Pr}_F[Y^n\in\mathcal A_n^{2k'}(V^n,S)] - o(1) \right]\\
&\stackrel{(f)}{\geq } &  \frac{\gamma_{\min}^{2k'}}{ B_n(2k')} (1-\epsilon- o(1)),
\label{eq:E3_Delta_sum}
\end{IEEEeqnarray}
where recall that $B_n(r)$ denotes the cardinality of the  $n$-dimensional $r$-ball defined in \eqref{eq:hammmin_bamm} and 
$\gamma_{\min}$ was defined in \eqref{eq:E3_gamma_min}. 
In the above, Inequality $(d)$ holds by simple counting arguments; Equality $(e)$ holds by standard typicality arguments, by 
 \eqref{eq:E3_low_weight}, and  because $F_U=Q_U$;  and Inequality~$(f)$ holds by \eqref{eq:E3_enlarged_auxiliary}.
%To see Equality $(e)$ more in detail, notice that for any $x^n\in\mathcal X^n$ and $y^n,\tilde y^n\in\mathcal Y^n$ satisfying $d_{\mathrm H}(y^n,\tilde y^n)\le r$: 
%\begin{IEEEeqnarray}{rCl}
%\frac{\Gamma_{Y|X}^{\otimes n}(\tilde y^n|x^n)}{\Gamma_{Y|X}^{\otimes n}(y^n|x^n)}
%&&=\prod_{i:\tilde y_i\neq y_i}\frac{\Gamma_{Y|X}(\tilde y_i|x_i)}{\Gamma_{Y|X}(y_i|x_i)} %\leq\gamma_{\min}^{-r},
%\label{eq:E3_gamma_ratio}
%\end{IEEEeqnarray}
%and  
%\begin{IEEEeqnarray}{rCl}
%\lefteqn{ {\Pr}_F[Y^n\in\mathcal A_n^{r}(V^n,S)]} \nonumber \\
%& \leq & \sum_{y^n\in \mathcal{Y}^n} \quad \sum_{\tilde{y}^n \in \mathbb{B}^r_n(y^n)} %{\Pr}_F[ Y^n=\tilde{y}^n, \tilde{y}^n\in\mathcal A_n^{r}(V^n,S)] \IEEEeqnarraynumspace\\
%&&\leq\gamma_{\min}^{-r}|B_n^r|\sum_{y^n\in \mathcal{Y}^n}  {\Pr}_F[ Y^n={y}^n, \tilde{y}^n\in\mathcal A_n^{r}(V^n,S)] \\
%&&=\gamma_{\min}^{-r}|B_n^r|  {\Pr}_F[ Y^n \in\mathcal A_n^{r}(V^n,S)] .
%\label{eq:E3_ball_bound}
%\end{IEEEeqnarray}

By  the standard Hamming-ball cardinality bound
(see, e.g., \cite[Proof of Lemma~5.1]{Csiszarbook}) and since $k'/n\to 0$ (see \eqref{eq:E3_kprime_limit}) we have that $\lim_{n\to \infty} \frac{1}{n} \log B_n(2k')=0$. Therefore   \eqref{eq:E3_Delta_sum} establishes \eqref{eq:f}, which concludes the proof.

\section{Conclusion}

We studied distributed binary hypothesis testing over discrete memoryless channels (DMC) under a covertness constraint imposed either under the null hypothesis alone or under both hypotheses. We characterized the optimal Stein-exponent in both scenarios and showed that it depends on the connectivity of the DMC to the legitimate receiver.

For partially-connected DMCs, both covertness requirements yield the Shalaby–Papamarcou Stein-exponent for a noiseless link with a sublinear number of communication bits. This exponent does not depend on the specific DMC transition probabilities. For fully-connected DMCs, the Stein-exponent also depends on the DMC law. Requiring covertness under both hypotheses introduces an additional bound and can reduce the optimal Stein-exponent. Nevertheless, covert communication can strictly improve detection performance compared with a test based only on the decision center’s local observations.

Under the stated assumptions, these Stein-exponents coincide with those obtained under the corresponding expected sublinear input-cost constraints. Thus, although covertness imposes a stronger restriction on communication, it causes no additional loss in the optimal Stein-exponent in these models. Notice that  the resulting Stein-exponents are independent of the prescribed Type-I error threshold. Moreover, while our converse bounds  allow for shared secret keys between the sensor and the decision center of arbitrary length, our achievability results do not require such a shared key, except when covertness is imposed under both hypotheses and the DMC is partially-connected. In this latter scenario a shared key of length that is slightly above the logarithm of the observation length suffices.  

Natural extensions of this work include the study of systems with multiple sensors or decision centers, as well as covertness constraints formulated in terms of total variation distance or other distance measures.
 \section*{Acknowledgments}
The work of the authors has been supported by the ERC under grant agreement 101125691.

\appendices  

\begin{appendices}

\section{Proof of Equation~\eqref{eq:TT}} \label{app:proof_lemma_div}

Let $\mathcal K\triangleq\{s\subseteq \{1,\ldots,n\}:|s|=m_n\}$.
For each $s\in\mathcal K$, define
\begin{IEEEeqnarray}{rCl}
P_{Z^n|S=s}(z^n)
&\triangleq&
\prod_{i\in s}\Gamma_{Z|X}(z_i|x_1)
\prod_{i\notin s}\Gamma_{Z|X}(z_i|0).
\label{eq:both_output_given_s}
\end{IEEEeqnarray}
Also define
\begin{IEEEeqnarray}{rCl}
\overline P_{Z^n}(z^n)
&\triangleq&\frac{1}{\binom n{m_n}}
\sum_{s\in\mathcal K}P_{Z^n|S=s}(z^n).
\label{eq:both_output_mixture}
\end{IEEEeqnarray}
Let further $\rho_{n, H} = \Pr[U^n\notin\mathcal T_\mu^{(n)}(P_U)|\mathcal H=H]$, and recall that for each $H\in\{0,1\}$,
\begin{IEEEeqnarray}{rCl}
\lefteqn{P_{Z^n|\mathcal H=H}(z^n)}\nonumber\\
%&=&\Pr[U^n\in\mathcal T_\mu^{(n)}(P_U)|\mathcal H=H]
%P_{Z^n|U^n\in\mathcal T_\mu^{(n)}(P_U),\mathcal H=H}(z^n)\nonumber\\
%&&+\Pr[U^n\notin\mathcal T_\mu^{(n)}(P_U)|\mathcal H=H]
%P_{Z^n|U^n\notin\mathcal T_\mu^{(n)}(P_U),%\mathcal H=H}(z^n)\nonumber\\
&=&(1-\rho_{n, H})\Gamma_{Z|X}^{\otimes n}(z^n|0^n)
+\rho_{n, H}\overline P_{Z^n}(z^n).
\label{eq:both_warden_mixture}
\end{IEEEeqnarray}

Then
\begin{IEEEeqnarray}{rCl}
\lefteqn{D\left(P_{Z^n|\mathcal H=H}\big\|\Gamma_{Z|X}^{\otimes n}(\cdot|0^n)\right)}\nonumber\\
&=&D\left((1-\rho_{n, H})\Gamma_{Z|X}^{\otimes n}(\cdot|0^n)+\rho_{n, H}\overline P_{Z^n}\big\|\Gamma_{Z|X}^{\otimes n}(\cdot|0^n)\right)\nonumber\\
&=&\sum_{z^n}P_{Z^n|\mathcal H=H}(z^n) \nonumber \\
&&\hspace{3mm} \cdot 
\log\left(1+\rho_{n, H}
\frac{\overline P_{Z^n}(z^n)-\Gamma_{Z|X}^{\otimes n}(z^n|0^n)}
{\Gamma_{Z|X}^{\otimes n}(z^n|0^n)}\right)\nonumber\\
&\stackrel{(a)}{\leq}&\rho_{n, H}\sum_{z^n}
P_{Z^n|\mathcal H=H}(z^n)
\frac{\overline P_{Z^n}(z^n)-\Gamma_{Z|X}^{\otimes n}(z^n|0^n)}
{\Gamma_{Z|X}^{\otimes n}(z^n|0^n)}\nonumber\\
&=&\rho_{n, H}\sum_{z^n}
\left[\overline P_{Z^n}(z^n)-\Gamma_{Z|X}^{\otimes n}(z^n|0^n)\right]\nonumber\\
&&+\rho_{n, H}^2\sum_{z^n}
\frac{\left[\overline P_{Z^n}(z^n)-\Gamma_{Z|X}^{\otimes n}(z^n|0^n)\right]^2}
{\Gamma_{Z|X}^{\otimes n}(z^n|0^n)}\nonumber\\
&=&\rho_{n, H}^2\sum_{z^n}
\frac{\left[\overline P_{Z^n}(z^n)-\Gamma_{Z|X}^{\otimes n}(z^n|0^n)\right]^2}
{\Gamma_{Z|X}^{\otimes n}(z^n|0^n)}\nonumber\\
&=&\rho_{n, H}^2\cdot \chi^2\left(\overline P_{Z^n}\big\|\Gamma_{Z|X}^{\otimes n}(\cdot|0^n)\right),
\label{eq:both_KL_chi2}
\end{IEEEeqnarray}
where $(a)$ follows from $\log(1+x)\leq x$.

\medskip
We now develop the $\chi^2$-divergence:
\begin{IEEEeqnarray}{rCl}
\lefteqn{\chi^2\left(\overline P_{Z^n}\big\|\Gamma_{Z|X}^{\otimes n}(\cdot|0^n)\right)}\nonumber\\
&=&\sum_{z^n}
\frac{\left[\overline P_{Z^n}(z^n)-\Gamma_{Z|X}^{\otimes n}(z^n|0^n)\right]^2}
{\Gamma_{Z|X}^{\otimes n}(z^n|0^n)}\nonumber\\
&=&\sum_{z^n}
\frac{\overline P_{Z^n}(z^n)^2}
{\Gamma_{Z|X}^{\otimes n}(z^n|0^n)}-1\nonumber\\
&=&\sum_{z^n}
\frac{\left(\frac{1}{\binom n{m_n}}
\sum_{s\in\mathcal K}P_{Z^n|S=s}(z^n)\right)^2}
{\Gamma_{Z|X}^{\otimes n}(z^n|0^n)}-1\nonumber\\
&=&\frac{1}{\binom n{m_n}^2}
\sum_{s_1,s_2\in\mathcal K}
\sum_{z^n}
\frac{P_{Z^n|S=s_1}(z^n)P_{Z^n|S=s_2}(z^n)}
{\Gamma_{Z|X}^{\otimes n}(z^n|0^n)}-1. \nonumber \\
\label{eq:both_chi2_expansion}
\end{IEEEeqnarray}

For any $s_1,s_2\in\mathcal K$, let
$s_1\triangle s_2\triangleq
(s_1\setminus s_2)\cup(s_2\setminus s_1)$. We have
\begin{IEEEeqnarray}{rCl}
\lefteqn{\frac{P_{Z^n|S=s_1}(z^n)P_{Z^n|S=s_2}(z^n)}
{\Gamma_{Z|X}^{\otimes n}(z^n|0^n)}}\nonumber\\
&=&\prod_{i\in s_1\cap s_2}
\frac{\Gamma_{Z|X}(z_i|x_1)^2}{\Gamma_{Z|X}(z_i|0)}
\prod_{i\in s_1\triangle s_2}\Gamma_{Z|X}(z_i|x_1)
\nonumber \\
&& \cdot \prod_{i\notin s_1\cup s_2}\Gamma_{Z|X}(z_i|0). 
\label{eq:both_product_expansion}
\end{IEEEeqnarray}
Thus
\begin{IEEEeqnarray}{rCl}
\lefteqn{\sum_{z^n}\frac{P_{Z^n|S=s_1}(z^n)P_{Z^n|S=s_2}(z^n)}
{\Gamma_{Z|X}^{\otimes n}(z^n|0^n)}}\nonumber\\
&=&\left(\sum_{z\in\mathcal Z}
\frac{\Gamma_{Z|X}(z|x_1)^2}{\Gamma_{Z|X}(z|0)}
\right)^{|s_1\cap s_2|}\nonumber\\
&=&\left(1+\chi^2(\Gamma_{Z|X=x_1}\|
\Gamma_{Z|X=0})\right)^{|s_1\cap s_2|}.
\label{eq:both_product_sum}
\end{IEEEeqnarray}
Substituting \eqref{eq:both_product_sum} into
\eqref{eq:both_chi2_expansion}, we obtain
\begin{IEEEeqnarray}{rCl}
\lefteqn{\chi^2\left(\overline P_{Z^n}
\big\|\Gamma_{Z|X}^{\otimes n}(\cdot|0^n)\right)}\nonumber\\
&=&\frac{1}{\binom n{m_n}^2}
\sum_{s_1,s_2\in\mathcal K}
\left(1+\chi^2(\Gamma_{Z|X=x_1}\|
\Gamma_{Z|X=0})\right)^{|s_1\cap s_2|}-1. \nonumber \\
\label{eq:both_chi2_intersection}
\end{IEEEeqnarray}

For any fixed $s_1\in\mathcal K$, the number of subsets
$s_2\in\mathcal K$ satisfying $|s_1\cap s_2|=j$ is
$\binom{m_n}{j}\binom{n-m_n}{m_n-j}$. Therefore,
\begin{IEEEeqnarray}{rCl}
\lefteqn{\sum_{s_2\in\mathcal K}
\left(1+\chi^2(\Gamma_{Z|X=x_1}\|\Gamma_{Z|X=0})\right)^{|s_1\cap s_2|}}\nonumber\\
&=&\sum_{j=0}^{m_n}
\binom{m_n}{j}\binom{n-m_n}{m_n-j}
\left(1+\chi^2(\Gamma_{Z|X=x_1}\|\Gamma_{Z|X=0})\right)^j. \nonumber \\
\label{eq:both_fixed_s1}
\end{IEEEeqnarray}
Substituting \eqref{eq:both_fixed_s1} into
\eqref{eq:both_chi2_intersection} yields
\begin{IEEEeqnarray}{rCl}
\lefteqn{\chi^2\left(\overline P_{Z^n|S}
\big\|\Gamma_{Z|X}^{\otimes n}(\cdot|0^n)\right)}\nonumber\\
&=&\sum_{j=0}^{m_n}
\frac{\binom{m_n}{j}\binom{n-m_n}{m_n-j}}
{\binom{n}{m_n}}
\left(1+\chi^2(\Gamma_{Z|X=x_1}\|\Gamma_{Z|X=0})\right)^j-1\nonumber\\
&\overset{(b)}{=}&\sum_{j=0}^{m_n}
\frac{\binom{m_n}{j}\binom{n-m_n}{m_n-j}}
{\binom{n}{m_n}}
\sum_{k=0}^{j}\binom jk
\left[\chi^2(\Gamma_{Z|X=x_1}\|\Gamma_{Z|X=0})\right]^k-1\nonumber\\
&=&\sum_{k=0}^{m_n}
\frac{\left[\chi^2(\Gamma_{Z|X=x_1}\|\Gamma_{Z|X=0})\right]^k}
{\binom{n}{m_n}} \nonumber \\
&&\hspace{3cm}\cdot
\sum_{j=k}^{m_n}
\binom{m_n}{j}\binom{n-m_n}{m_n-j}\binom jk-1\nonumber\\
&\overset{(c)}{=}&\sum_{k=0}^{m_n}
\frac{\left[\chi^2(\Gamma_{Z|X=x_1}\|\Gamma_{Z|X=0})\right]^k}
{\binom{n}{m_n}} \nonumber \\
&&\hspace{3cm}\cdot\sum_{j=k}^{m_n} \binom{n-m_n}{m_n-j}
\binom{m_n}{k}\binom{m_n-k}{j-k}-1\nonumber\\
&\overset{(d)}{=}&\sum_{k=0}^{m_n}
\frac{\left[\chi^2(\Gamma_{Z|X=x_1}\|\Gamma_{Z|X=0})\right]^k}
{\binom{n}{m_n}} \nonumber \\
&&\hspace{2cm}\cdot\sum_{r=0}^{m_n-k} 
\binom{m_n}{k} \binom{m_n-k}{r} \binom{n-m_n}{m_n-k-r}-1\nonumber\\
&\overset{(e)}{=}&\sum_{k=0}^{m_n}
\binom{m_n}{k}
\frac{\binom{n-k}{m_n-k}}{\binom{n}{m_n}}
\left[\chi^2(\Gamma_{Z|X=x_1}\|\Gamma_{Z|X=0})\right]^k-1\nonumber\\
&\overset{(f)}{\leq}&\sum_{k=0}^{m_n}
\binom{m_n}{k}
\left[\frac{m_n}{n}
\chi^2(\Gamma_{Z|X=x_1}\|\Gamma_{Z|X=0})\right]^k-1\nonumber\\
&=&\left(1+\frac{m_n}{n}
\chi^2(\Gamma_{Z|X=x_1}\|\Gamma_{Z|X=0})\right)^{m_n}-1,
\label{eq:both_combinatorial_bound}
\end{IEEEeqnarray}
where (b) holds by the binomial theorem; (c)  by the identity
$\binom{m_n}{j}\binom{j}{k}
=\binom{m_n}{k}\binom{m_n-k}{j-k}$;
(d)  by the change of variable $r=j-k$; (e)  by Vandermonde's identity
\begin{IEEEeqnarray}{rCl}
\sum_{r=0}^{m_n-k}
\binom{m_n-k}{r}
\binom{n-m_n}{m_n-k-r}
&=&\binom{n-k}{m_n-k};
\end{IEEEeqnarray}
and (f) by
\begin{IEEEeqnarray}{rCl}
\frac{\binom{n-k}{m_n-k}}{\binom{n}{m_n}}
&=&\prod_{i=0}^{k-1}\frac{m_n-i}{n-i} \leq \left(\frac{m_n}{n}\right)^k.
\end{IEEEeqnarray}

Thus we obtain
\begin{IEEEeqnarray}{rCl}
\lefteqn{\chi^2\left(\overline P_{Z^n|S}
\big\|\Gamma_{Z|X}^{\otimes n}(\cdot|0^n)\right)}\nonumber\\
&\leq&\left(1+\frac{m_n}{n}
\chi^2(\Gamma_{Z|X=x_1}\|\Gamma_{Z|X=0})\right)^{m_n}-1.
\label{eq:both_chi2_final}
\end{IEEEeqnarray}
Combining \eqref{eq:both_KL_chi2} and
\eqref{eq:both_chi2_final} yields
\begin{IEEEeqnarray}{rCl}
\lefteqn{D\left(P_{Z^n|\mathcal H=H}
\big\|\Gamma_{Z|X}^{\otimes n}(\cdot|0^n)\right)}\nonumber\\
&\leq&\rho_{n, H}^2
\left[\left(1+\frac{m_n}{n}
\chi^2(\Gamma_{Z|X=x_1}\|\Gamma_{Z|X=0})\right)^{m_n}-1\right].\IEEEeqnarraynumspace
\label{eq:both_covertness_bound_proved}
\end{IEEEeqnarray}

This completes the proof of  Equation~\eqref{eq:TT}.

\section{Proof of Lemma~\ref{lem:lemma_low_weight}\label{app:proof_low_weight}}
Start by noticing that under $\mathcal{H}=0$: 
\begin{IEEEeqnarray}{rCl}
	\lefteqn{D( P_{Z^n|\mathcal{H}=0} \| \Gamma_{Z|X}^{\otimes n}(\cdot|0^n))} \nonumber \\
	&=& - H(Z^n) +\mathbb{E}_{Z^n} \left[ \log \left( \frac{1}{\Gamma_{Z|X}^{\otimes n}(Z^n | 0^n)} \right ) \right ] \\
%	&=& - \sum_{i=1}^n H(Z_i \mid Z^{i-1}) + \E_{Z_i} \left[ \log \left( \frac{1}{\Gamma_{Z|X}(Z_i | 0)} \right ) \right ] \\
	&\overset{(a)}{\geq}& - \sum_{i=1}^n H(Z_i) + \mathbb{E}_{Z_i} \left[ \log \left( \frac{1}{\Gamma_{Z|X}(Z_i | 0)} \right ) \right ] \\
    &=&\sum_{i=1}^n D\left(
P_{Z_i|\mathcal{H}=0} \middle\|
\Gamma_{Z|X}(\cdot \mid 0) \right)\\
	&\overset{(b)}{=}& \sum_{i=1}^n D \Big( \sum_{x \in \mathcal{X}\setminus\{0\}} b_{n,i}^{0}(x) a_{n,i}^0 \Gamma_{Z|X}(\cdot \mid x) \nonumber \\
	&& \hspace{0.5cm} + \  (1-a_{n,i}^0) \Gamma_{Z|X}(\cdot \mid 0) \| \hspace{0.1cm} \Gamma_{Z|X}(\cdot \mid 0) \Big )  ,\label{eq:to_prove_vanishing_alpha_n_i_0}
\end{IEEEeqnarray}
where $(a)$ holds because conditioning reduces entropy; and $(b)$ because: 
\begin{IEEEeqnarray}{rCl}
\lefteqn{P_{Z_i|\mathcal{H}=0}(z)} \nonumber \\
&=& \Pr[Z_i = z \mid X_i = 0, \mathcal{H}=0]\Pr[X_i=0 \mid \mathcal{H}=0]  \nonumber\\
   && + \ \Pr[Z_i = z \mid X_i \neq 0, \mathcal{H}=0]\Pr[X_i \neq 0 \mid \mathcal{H}=0] \IEEEeqnarraynumspace \\
&= &(1-\Pr[X_i \neq 0 \mid \mathcal{H}=0]) \nonumber \\
&& \cdot \ \Pr(Z_i = z \mid X_i = 0, \mathcal{H}=0)  + \Pr[X_i \neq 0 \mid \mathcal{H}=0]\nonumber \\
   && \cdot \ 
   \sum_{x \in \mathcal{X}\setminus\{0\}} \Pr(Z_i = z \mid X_i = x, \mathcal{H}=0) \nonumber \\
   && \cdot \ \Pr(X_i = x \mid X_i \neq 0, \mathcal{H}=0) \\
&=& (1-a_{n,i}^0)\,\Gamma_{Z|X}(z \mid 0)
  \ \  \nonumber \\
   && + \ a_{n,i}^0 \sum_{x \in \mathcal{X}\setminus\{0\}} b_{n,i}^0(x) \,\Gamma_{Z|X}(z \mid x).
\end{IEEEeqnarray}

By the covertness constraint \eqref{eq:cov} for $\mathcal{H}=0$ and the nonnegativity of Kullback-Leibler divergence, we deduce that  each summand on the right-hand side of \eqref{eq:to_prove_vanishing_alpha_n_i_0} must vanish, and thus

\begin{equation}
\lim_{n\to \infty} a_{n,i}^{0} = 0, \quad  i\in\{1,\ldots, n\},
\end{equation}
and 
\begin{equation}\label{eq:alpha0}
\lim_{n\to \infty} \frac{1}{n} \sum_{i=1}^n a_{n,i}^{0} =0,
\end{equation}
which establishes Limit  $\lim_{n\to\infty} a_n^{0}$, i.e.,  \eqref{eq:van_a0}, in the lemma. 

%Let $\mathcal{A}$ be the input alphabet such that $0 \in \mathcal{A}$. Without loss of generality, consider also that $card(\mathcal{A}) = A+1 < \infty$.
  Notice next that by Markov's inequality:
\begin{IEEEeqnarray}{rCl}
 \sum_{i=1}^n a_{n,i}^{0} &= & \mathbb{E}[ w_{\textnormal{H}}(X^n) | \mathcal{H}=0] \\
 & \geq  &  \bar a_n^{0} \cdot n \cdot  \Pr[X^n \notin \tilde{\mathcal{X}}_{0}^n | \mathcal{H}=0].
 \end{IEEEeqnarray}
By \eqref{eq:eps_n} this further  implies 
\begin{IEEEeqnarray}{rCl}
\Pr\left[X^n \notin \tilde{\mathcal{X}}_0^n | \mathcal{H}=0\right] \leq \frac{ \frac{1}{n}  \sum_{i=1}^n a_{n,i}^{0}}{\bar a_n^{0}} 
& = & \sqrt{ \frac{1}{n}  \sum_{i=1}^n a_{n,i}^{0}}. \IEEEeqnarraynumspace
\end{IEEEeqnarray} 
By \eqref{eq:alpha0} this proves \eqref{eq:high_prob}.

Similar steps can be used to show that \eqref{eq:high_prob_both}  holds when  the covertness constraint is imposed under both $\mathcal{H}=0$ and $\mathcal{H}=1$.

To see that the size of $\tilde{\mathcal{X}}_0^n$ does not grow exponentially in $n$, we notice that this set  can  be described as the union over all  type-classes (i.e., sets of sequences  with same type) for types that assign frequency larger or equal to  $1-\bar a_n^0$ to the 0 symbol. Since  the type-class for type $\boldsymbol{\pi}$ is of size at most $2^{n H(\boldsymbol{\pi})}$ and since the number of type-classes is bounded by $(n+1)^{|\mathcal{X}|}$, we have: 
\begin{IEEEeqnarray}{rCl}
 |\tilde{\mathcal{X}}_0^n| &\leq&  (n+1)^{|\mathcal{X}|} 2^{n \max_{\boldsymbol{\pi}} H( \boldsymbol{\pi})},
\end{IEEEeqnarray} 
where the maximum is over all types $\boldsymbol{\pi}$ with $\boldsymbol{\pi}(0)\geq  1-\bar a_n^0$. 
Since $\bar a_n^0$ vanishes as $n\to \infty$ and by the continuity of the entropy functional, we  obtain 
\begin{IEEEeqnarray}{rCl}\label{eq:cardinality_limit}
\lim_{n\to \infty} \frac{1}{n} \log \left| \tilde{\mathcal{X}}_0^n \right|&\leq & \lim_{n\to \infty} \left[ \frac{|\mathcal{X}|}{n} \log (n+1) + \max_{ \substack{\boldsymbol{\pi} \colon \\ \boldsymbol{\pi}(0) \geq 1-\bar a_n^0}}  H( \boldsymbol{\pi}) \right] \nonumber \\
&=&0,
\end{IEEEeqnarray} 
proving \eqref{eq:sizeX}. 

Similar arguments can be used to prove the second part of  the lemma.

\section{Proof of Lemma~\ref{lem:lemma_ps}}\label{app:proof_lemma_ps}
Define the sets 
\begin{IEEEeqnarray}{rCl}
\bar{\mathcal{C}}_{\bar{s}} &\triangleq &\{ u^n \in \mathcal{U}^n \colon f^{(n)}(u^n, \bar{s})= \bar{x}^n\}\\
\bar{\mathcal{D}}_{\bar{s}} &\triangleq &\{ v^n \in \mathcal{V}^n \colon g^{(n)}(v^n,\bar{x}^n, \bar{s})= 0\}.
\end{IEEEeqnarray}

Note that 
\begin{IEEEeqnarray}{rCl}
\lefteqn{P_{UV}^{\otimes n} (\bar{\mathcal{C}}_{\bar{s}} \times \bar{\mathcal{D}}_{\bar{s}}) } \nonumber \\
&=&\hspace{-0.6cm}\sum_{\substack{v^n, u^n \colon \\
f^{(n)}(u^n, \bar{s})= \bar{x}^n \\
g^{(n)}(v^n,\bar{x}^n, \bar{s})= 0}} \hspace{-0.5cm}  \Pr[U^n=u^n, V^n = v^n | \mathcal{H}=0, S=\bar{s}] 
\\
&=&  \Pr[\mathcal{\hat{H}}=0, X^n = \bar{x}^n | \mathcal{H}=0, S=\bar{s}], 
\end{IEEEeqnarray}
which by assumption exceeds $e^{-n \phi_n}$. %$  \frac{1- \epsilon-\eta}{c|\tilde{\mathcal{X}}^n|}$.

%By definition of $\bar{S}_{\bar{x}^n}$, 
%P_{UV}^{\otimes n} (\bar{\mathcal{C}}_{\bar{s}} \times \bar{\mathcal{D}}_{\bar{s}}) \geq   \frac{1- \epsilon-\eta}{c|\tilde{\mathcal{X}}^n|}, \ \ \   \forall \bar{s} \in \bar{S}_{\bar{x}^n}
%\end{equation}
Pick now an arbitrary $\tilde{P}_{UV}$ with marginals $\tilde{P}_U=P_{U}$ and $\tilde{P}_{V}=P_V$ and notice 
\begin{subequations}\label{eq:non_negligible}
\begin{IEEEeqnarray}{rCl}
\tilde{P}_U^{\otimes n}(\bar{\mathcal{C}}_{\bar{s}}) &= & P_U^{\otimes n}(\bar{\mathcal{C}}_{\bar{s}})\geq  e^{-n \phi_n}
\end{IEEEeqnarray}
and 
\begin{IEEEeqnarray}{rCl}
\tilde{P}_V^{\otimes n}(\bar{\mathcal{D}}_{\bar{s}}) &= & P_V^{\otimes n}(\bar{\mathcal{D}}_{\bar{s}}) \geq e^{-n \phi_n}.
\end{IEEEeqnarray}
\end{subequations}

In the following we shall slightly blow up (enlarge) the sets $\bar{\mathcal{C}}_{\bar{s}}$ and $\bar{\mathcal{D}}_{\bar{s}}$ to obtain $\hat{\mathcal{C}}_{\bar{s}}$ and $\hat{\mathcal{D}}_{\bar{s}}$, and then show that these enlarged sets contain a large portion of the typical set $\mathcal{T}_{\mu_n}^{(n)}(\tilde{P}_{UV})$.  

Notice that by the  blowing-up lemma \cite[Remark on p. 446]{MartonBU}, there exists a suitable sequence $\{\ell_n\}_{n\geq 1}$ (depending on $\phi_n$)  satisfying $\lim_{n\to\infty}\ell_n/\sqrt{n}=\infty$ and $\lim_{n\to\infty}\ell_n/n=0$, and   so that the blown up regions
\begin{IEEEeqnarray}{rCl}	\hat{\mathcal{C}}_{\bar{s}} &\triangleq& \left\{ \tilde{u}^n\colon \exists u^n\in\bar{\mathcal{C}}_{\bar{s}} \;\; \textnormal{s.t.}\;\;d_{\text{H}}(\tilde{u}^n,u^n)\leq \ell_n \right\} \\
\hat{\mathcal{D}}_{\bar{s}} &\triangleq& \left\{ \tilde{v}^n\colon \exists v^n\in\bar{\mathcal{D}}_{\bar{s}} \;\; \textnormal{s.t.}\;\;d_{\text{H}}(\tilde{v}^n,v^n)\leq \ell_n \right\}
\end{IEEEeqnarray}
satisfy
\begin{subequations}\label{eq:lambda_n}
\begin{IEEEeqnarray}{rCl}
\tilde{P}_U^{\otimes n}(\hat{\mathcal{C}}_{\bar{s}}) &\geq&	1 -\lambda_n, \\ 
\tilde{P}_V^{\otimes n}(\hat{\mathcal{D}}_{\bar{s}}) &\geq&	1 -\lambda_n, 
\end{IEEEeqnarray}
\end{subequations}
for some  sequence $\lambda_n$ that tends to 0 as $n\to \infty$. Then, by the union bound: 
\begin{equation} 
\tilde{P}_{UV}^{\otimes n} (\hat{\mathcal{C}}_{\bar{s}} \times \hat{\mathcal{D}}_{\bar{s}}) \geq 1- 2\lambda_n. 
\end{equation} 

Moreover, standard counting arguments yield 
%Continue with the following considerations:
\begin{IEEEeqnarray}{rCl}
&&\hspace{-0.8cm}\Pr\left[ \hat{\mathcal{H}}=0, X^n=\bar{x
}^n | \mathcal{H}=1, S=\bar{s}\right]  \nonumber\\
&=&Q_{UV}^{\otimes n} \left( \bar{\mathcal{C}}_{s} \times \bar{ \mathcal{D}}_{s}\right) 
 \\
&\geq& 
Q_{UV}^{\otimes n}
\left(
\hat{\mathcal C}_{s}
\times
\hat{\mathcal D}_{s}
\right)
\cdot
\frac{q_{\min}^{2\ell_n}}{B_n(2\ell_n)},
\label{eq:CD}
\label{eq:CD}
\end{IEEEeqnarray}
where 
\begin{IEEEeqnarray}{rCl}
q_{\min}
&\triangleq&
\min_{(u,v)\in\mathcal U\times\mathcal V}
Q_{UV}(u,v)>0,
\end{IEEEeqnarray}
and $B_n(r)$ denotes the cardinality of the  $n$-dimensional $r$-ball defined in \eqref{eq:hammmin_bamm}.
%\begin{equation}
%\xi_n \triangleq H_{\textnormal{b}}(\ell_n/n)+ \frac{\ell_n}{n} \log (| \mathcal{U}||\mathcal{V}|) - \frac{\ell_n}{n} \log \min_{(u,v)} \underbrace{Q_{UV}(u,v)}_{>0}.
%\end{equation}

The desired inequality is finally obtained from \eqref{eq:CD} and the following 
\begin{equation}
Q_{UV}^{\otimes n}(\hat C_{\bar s}\times\hat D_{\bar s})
\ge
e^{-\frac{nD(\widetilde P_{UV}\|Q_{UV})+\log 2}{1-2\lambda_n}},
\label{eq:appC-direct-change}
\end{equation}
which we prove next. Define
\begin{IEEEeqnarray}{rCl}
p_n & \triangleq &\tilde{P}_{UV}^{\otimes n}(\hat C_{\bar s}\times\hat D_{\bar s})\\
q_n&\triangleq &Q_{UV}^{\otimes n}(\hat C_{\bar s}\times\hat D_{\bar s}).
\end{IEEEeqnarray}
Then, by the data-processing inequality 
\begin{IEEEeqnarray}{rCl}
\lefteqn{nD(\widetilde P_{UV}\|Q_{UV})}\\
&=&D_(\widetilde P_{UV}^{\otimes n}\|Q_{UV}^{\otimes n})\nonumber\\
&\ge &D_b(p_n\|q_n)\nonumber\\
&=&-H_b(p_n)+p_n\log\frac1{q_n}
 +(1-p_n)\log\frac1{1-q_n}\nonumber\\
&\ge &-\log 2+(1-2\lambda_n)\log\frac1{q_n},
\label{eq:exp_bound}
\end{IEEEeqnarray}
which establishes \eqref{eq:exp_bound}.

Combining \eqref{eq:CD} and \eqref{eq:appC-direct-change} yields the desired result.

\end{appendices}

\bibliographystyle{ieeetr}
\bibliography{references}

\end{document}